\documentclass[preprint]{aastex}

\newcommand       \kpc          {\,{\rm kpc}}

\newcommand       \mum          {\,{\rm \mu m}}
\newcommand       \Ks           {{\rm K_{s}}}
\newcommand       \J            {{\rm J}}
\newcommand       \HH           {{\rm H}}
\newcommand       \K            {{\rm K}}
\newcommand       \simali       {\,{\sim}}
\newcommand       \magni        {\,{\rm mag}}
\newcommand       \tausil       {\Delta\tau_{9.7\mum}}
\newcommand       \tauahc       {\Delta\tau_{3.4\mum}}

\shorttitle{Variation of Mid-Infrared Extinction}
\shortauthors{Gao, Jiang \& Li}

\begin{document}

\title{Mid-Infrared Extinction and its Variation with Galactic Longitude}


\author{Jian Gao\altaffilmark{1,2},
        B.~W. Jiang\altaffilmark{1}
        and Aigen Li\altaffilmark{2}}

\altaffiltext{1}{Department of Astronomy, Beijing Normal University,
Beijing 100875, China; {\sf jiangao@bnu.edu.cn, bjiang@bnu.edu.cn}}
 \altaffiltext{2}{Department of Physics and
Astronomy, University of Missouri, Columbia, MO 65211, USA.;
{\sf lia@missouri.edu}}

\begin{abstract}
Based on the data obtained from the \textit{Spitzer}/GLIPMSE Legacy
Program and the 2MASS project, we derive the extinction in the four
IRAC bands, [3.6], [4.5], [5.8] and [8.0]$\mum$, relative to the
2MASS $\Ks$ band (at 2.16$\mum$) for 131 GLIPMSE fields along the
Galactic plane within $|l|\leq65^{\rm o}$, using red giants and red
clump giants as tracers. As a whole, the mean extinction in the IRAC
bands (normalized to the 2MASS $\Ks$ band),
$A_{[3.6]}/A_\Ks\approx0.63\pm0.01$,
$A_{[4.5]}/A_\Ks\approx0.57\pm0.03$,
$A_{[5.8]}/A_\Ks\approx0.49\pm0.03$,
$A_{[8.0]}/A_\Ks\approx0.55\pm0.03$, exhibits little variation with
wavelength (i.e. the extinction is somewhat flat or gray). This is
consistent with previous studies and agrees with that predicted from
the standard interstellar grain model for $R_V=5.5$ by
\citet{Weingartner01}. As far as individual sightline is
concerned, however, the wavelength dependence of the mid-infrared
interstellar extinction $A_{\lambda}/A_\Ks$ varies from one
sightline to another, suggesting that there may not exist a
``universal'' IR extinction law. We, for the first time,
demonstrate the existence of systematic variations of extinction
with Galactic longitude which appears to correlate with the
locations of spiral arms as well as with the variation of the far
infrared luminosity of interstellar dust.
\end{abstract}


\section{Introduction}
With the development of space infrared (IR) astronomy, the precise
determination of IR extinction becomes urgent in order to recover
the intrinsic colors and spectral energy distributions (SEDs) of
heavily obscured sources. There have been various attempts to
measure the IR extinction based on the \emph{Infrared Space
Observatory}(\emph{ISO}) and \emph{Spitzer Space Telescope} since
\citet{Lutz96} obtained the mid-IR extinction from several hydrogen
recombination lines and demonstrated the absence
of the model-predicted pronounced minimum around 7$\mum$.
This was supported by
\citet{Jiang03, Jiang06} based on the ISOGAL database
\citep{Omont99}, and by \citet{Indebetouw05} based on the data from
the \emph{Spitzer} Galactic Legacy Infrared Midplane Survey
Extraordinaire (GLIMPSE) Legacy Program \citep{Benjamin03}. All
these results roughly agree with the extinction predicted by the
standard interstellar grain model for $R_V=5.5$ of
\citet{Weingartner01} and \citet{Draine03}.\footnote{%
  $R_V\equiv A_V/E(B-V)$ is the total-to-selective extinction ratio,
  where $E(B-V)\equiv A_B-A_V$, the color excess, is the difference
  between the extinction in $B$ and $V$ bands.
  }
However, so far only a few wave bands have been investigated and the
sky coverage is also limited. No consensus has been reached yet
regarding the interstellar extinction at $\simali$5--8$\mum$.

Recent progress in the IR extinction measurements is made toward
star-forming regions mainly based on the \emph{Spitzer}
observations. Thanks to the high sensitivity of \emph{Spitzer}, deep
photometry is now possible and objects that suffer severe extinction
are now reachable. \citet{Flaherty07} studied five nearby
star-forming regions at mid-IR wavelengths
(3.6$\mum$, 4.5$\mum$, 5.8$\mum$ and 8.0$\mum$
from the InfraRed Array Camera [IRAC],\footnote{%
  The effective wavelengths of the four IRAC bands are actually
  3.545$\mum$, 4.442$\mum$, 5.675$\mum$ and 7.760$\mum$, respectively.
  }
and 24$\mum$ from the Multiband Imaging Photometer [MIPS]).
They confirmed a relatively flat extinction
curve at $\simali$4--8$\mum$. \citet{Roman07} studied a star-forming
dense cloud core located in the Pipe Nebula, and found that the IR
extinction in the IRAC bands of that region also agrees with
the $R_{V}=5.5$ model curve and indicates a dust size distribution
favoring larger sizes.

Although both the \emph{ISO} and \emph{Spitzer} measurements agree
with each other in that the $\simali$5--8$\mum$ extinction is
relatively flat and lacks the model-predicted minimum around 7$\mum$
of \citet{Draine89}, there do exist differences among various
measurements made for different sightlines. In their sample of five
sightlines toward star-forming regions, \citet{Flaherty07} found a
clear difference between one sightline and the other four sightlines
(see their Table 3). They derived higher $A_{\lambda}/A_\Ks$ ratios
and a flatter wavelength dependence than that of
\citet{Indebetouw05} for the same sightline toward $l=284^{\circ}$
in the Galactic Plane.

From $>$\,200 fields observed in the ISOGAL survey, \citet{Jiang06}
analyzed the extinction at 7$\mum$ and 15$\mum$ along $\simali$120
directions. They found marginal variation of the extinction at
7$\mum$. It is commonly believed that, with the parameter $R_V$
increasing in denser regions, the variation of the ultraviolet (UV)
and visual extinction with wavelength becomes flatter than that of
the diffuse interstellar medium (ISM) which is characterized with a
lower $R_V$ \citep{Cardelli89}, while the near-IR extinction seems
to be ``universal", with little variation among different sightlines
\citep{Draine89}. However, \citet{Nishiyama06a, Nishiyama09}
recently argued against such a universal near-IR extinction. In
addition, \citet{Fitzpatrick04} argued that the IR-through-UV
Galactic extinction curves should not be considered as a simple
one-parameter family, whether characterized by $R_V$ as suggested by
\citet{Cardelli89} or any other parameters.

\citet{Whittet77} presented observational evidence for a small but
appreciable variation in $R_V$ with Galactic longitude. He suggested
that the most likely explanation for this is a variation in the mean
size of the dust in the local spiral arm. However, unfortunately
only a few data points were used in that work and therefore no
systematic variation of the extinction with Galactic longitude was
reported. \citet{Jiang06} obtained the extinction around 7$\mum$ for
129 different sightlines and no clear variation with Galactic
longitude was found, although the extinction ratio $A_{[7]}/A_\Ks$
does appear to exhibit a tendency of decreasing toward the Galactic
center where $|l|<2^{\circ}$\citep{Jiang06}. The
GLIMPSE Legacy Program surveyed the Galactic plane, with a large
area coverage ($|l|\leq65^{\circ}$) and a detection limit
of $\simali$15.5--13.0$\magni$ from 3.6 to 8.0$\mum$ \citep{GQA}.
It provides an opportunity to explore the systematic variation of
interstellar extinction in the IR with Galactic longitude.

In this work, we explore whether the mid-IR extinction varies among
sightlines and how it varies in different interstellar environments
based on the \emph{Spitzer}/GLIMPSE database. In \S\ref{DATA}, the
GLIMPSE data used in this work is briefly described. \S\ref{method}
presents the method adopted to derive the extinction. In
\S\ref{tracer} we discuss the selection of two different types of
tracers (i.e. red giants and red clump giants). \S\ref{results}
reports the resulting extinction ratios $A_{\lambda}/A_\Ks$ and the
mean extinction from the total 131 GLIMPSE fields. Also discussed in
\S\ref{results} are the comparison of the extinction derived here
with previous studies performed by \citet{Indebetouw05} and
\citet{Flaherty07}, and the longitudinal variation of the extinction
ratios $A_{\lambda}/A_\Ks$ as well as its relation with the
Galactic spiral arms and the distribution of interstellar dust. In
\S6 we summarize our major conclusions.

\section{Data: GLIMPSE and 2MASS}\label{DATA}
The data used in this work is obtained by the GLIMPSE group. GLIMPSE
is a \emph{Spitzer} Legacy Program to carry out an IR survey of the
inner Galactic plane using the IRAC camera on board the
\emph{Spitzer} Space Telescope \citep{Benjamin03}. It spans three
cycles, GLIMPSE, GLIMPSE-II and GLIMPSE-3D (see
\citealt{Churchwell09}). These GLIMPSE programs observed a large
part of the Galactic disk, including various sightlines along the
Galactic plane and providing an opportunity to investigate whether
the mid-IR extinction varies from one sightline to another.

The GLIMPSE and GLIMPSE-II enhanced data
have now been released.\footnote{%
  Data are available on
  \sf{http://www.astro.wisc.edu/sirtf/glimpsedata.html}
  }
These products consist of the highly reliable Point Source Catalogs
(GLMC), the more complete Point Source Archives (GLMA) and mosaic
images of the survey areas. In this study we will use the GLMC
catalogs since the sources in these catalogs were selected requiring
the reliability to exceed $\simali$99.5\% \citep{Churchwell09}.
Moreover, in the enhanced GLIMPSE version 2.0 catalogs and
GLIMPSE-II version 1.0 catalogs, the point sources were band-merged
(cross-identified) with the 2MASS Point Source Catalog (see
\citealt{Cutri03}). They provide both magnitudes and fluxes in the
four IRAC bands and three 2MASS bands.

Based on the original GLIMPSE and GLIMPSE-II Catalog files, we
divided the Galactic plane into 131 fields.\footnote{%
  The fields overlapped by GLIMPSE and GLIMPSE-II
          are combined into one field,
          i.e. the fields of $l=9.0^{\circ}$ to $10^{\circ}$
          and $l=350^{\circ}$ to $351^{\circ}$.
          Meanwhile, because of the insufficiency of sources,
          the two fields from
          $l=65^{\circ}$ to $65.3^{\circ}$ and $l=294.8^{\circ}$
          to $295^{\circ}$ are incorporated into the fields
          $l=64^{\circ}$ to $65^{\circ}$ and $l=295^{\circ}$
          to $296^{\circ}$, respectively.
          }
In this work, only the sources with a signal-to-noise ratio
S/N\,$\geq$\,5 in all three 2MASS bands and four IRAC bands are
taken into account. More details about sample selection will be
described in \S\ref{tracer}. Benefiting from the numerous detections
by \emph{Spitzer}/IRAC, the number of sources with S/N\,$\geq$\,5 in
a single sky field exceeds 10,000 for almost all the GLIMPSE fields.
The only exception is the GLIMPSE Observation Strategy Validation
(OSV) field, with $l = 283.8^{\circ}$ to $284.6^{\circ}$ and
$|b|<1^{\circ}$, where the number of sources with S/N\,$\geq$\,5 is
9625.

\section{Method}\label{method}
\subsection{Color-Excess Method}
The determination of dust extinction is most commonly made by
comparing the flux densities of extincted and un-extincted
pairs of stars of the same spectral type \citep{Draine03}.
\citet{Lutz99} used the H recombination lines detected between
2.5--9$\mum$ to probe the extinction law in this wavelength
range of the Galactic center based on a comparison of
the observed line fluxes with that expected from
the standard Case B recombination. Here we adopt the
``color-excess'' method to obtain the extinction. This method
calculates the ratio of two color excesses which can be expressed as
following
\begin{equation}\label{slope}
k_{x}\equiv\frac{E(\lambda_{r}-\lambda_{x})}{E(\lambda_{c}-\lambda_{r})}
= \frac{(\lambda_{r}-\lambda_{x})_{\rm
observed}-(\lambda_{r}-\lambda_{x})_{\rm
intrinsic}}{(\lambda_{c}-\lambda_{r})_{\rm
observed}-(\lambda_{c}-\lambda_{r})_{\rm intrinsic}}
=\frac{A_{r}-A_{x}}{A_{c}-A_{r}}
\end{equation}
where $\lambda_{x}$ is the magnitude in the band $x$ of interest;
$\lambda_{r}$ is the magnitude in the reference band $r$ (which is
usually taken to be the K or $\Ks$ band); $\lambda_{c}$ is magnitude
in the comparison band $c$ (which is usually taken to be the J or H
band). Therefore, the extinction ratio of the $x$ band to the $r$ reference band is
\begin{equation}\label{ext}
A_x/A_r = 1 + k_{x}\left(1 - A_c/A_r\right) ~~.
\end{equation}
Note that $A_{c}/A_{r}$ is always greater than 1 (no matter whether
the J band or the H band is chosen as the comparison band) since it
is generally true that $A_{\rm J}, A_{\rm H} > A_\K, A_{\Ks}$.
Therefore $A_{x}/A_{r}$ increases with the decreasing of $k_{x}$.

The ``color-excess'' method is widely applied to photometric data
and can probe deeper than the spectrum-pair method. Most of the IR
extinction determination studies are performed using this method.
In the color-excess method, a group of sources that
have the same intrinsic color indices
$(\lambda_{r}-\lambda_{x})_{\rm intrinsic}$ and
$(\lambda_{c}-\lambda_{r})_{\rm intrinsic}$ are chosen, $k_{x}$ is
simply the slope of the line that linearly fits the observed color
indices $(\lambda_{r}-\lambda_{x})_{\rm observed}$ and
$(\lambda_{c}-\lambda_{r})_{\rm observed}$ \citep{Jiang03}. This is
a statistical method as it makes use of a large number of sources
and reduces the risk of depending on any individual objects with
large uncertainties in the determination of their intrinsic color
indices. Meanwhile, it is essential for the accuracy of the method
to have a homogeneous sample [i.e. with very small scatter
in the color indices $(\lambda_{r}-\lambda_{x})_{\rm intrinsic}$
and $(\lambda_{c}-\lambda_{r})_{\rm intrinsic}$].

\subsection{Comparison Bands: $\J$ and $\HH$}
From equations (\ref{slope}) and (\ref{ext}), it can be seen that
the determination of $A_x/A_r$, the ratio of the $x$-band extinction
to the extinction of the reference band $r$, requires the knowledge
of $A_c/A_r$, the ratio of extinction at the comparison band $c$
to that at the reference band $r$. In previous studies, both
the $\J$ band \citep{Indebetouw05,Jiang03,Jiang06,Roman07} and the
$\HH$ band \citep{Flaherty07} have been used as the comparison
band. The advantage of taking the $\J$ band as the comparison band
is that $E(\J-\Ks)$ is more sensitive to the extinction as
$E(\J-\Ks)$ is about twice $E(\HH-\Ks)$ for the same value
of $A_\Ks$ (the extinction at the $\Ks$ band). The advantage of
choosing the $\HH$ band as the comparison band is that there are
more red giants detected in the $\HH$ band than in the $\J$ band.
This would be particularly important when the sample size is small
which could influence the statistics. But for the sky fields studied
here, the sample size is not a problem thanks to the sensitivity
of \emph{Spitzer}/IRAC. In this work, we therefore adopt
the $\J$ band as the comparison band. The $\J$ band was also
selected as the comparison band in our previous studies of the
ISOGAL fields \citep{Jiang03,Jiang06}. But in order to compare with
the results of \citet{Flaherty07}, we also calculate $A_x/A_r$ with
the $\HH$ band taken as the comparison band.

We take $A_\J/A_\Ks=2.52$ (with $\J$ as the comparison band) or
$A_\HH/A_\Ks=1.56$ (with $\HH$ as the comparison band) as derived by
\citet{Rieke85} for sightlines toward the GC.\footnote{%
  \citet{Indebetouw05} estimated $A_\J/A_\Ks\approx2.5\pm0.2$
  and $A_\HH/A_\Ks\approx1.55\pm0.1$ for the $l=42^{\circ}$
  and $284^{\circ}$ sightlines in the Galactic Plane
  which are very close to that of \citet{Rieke85}.
  However, \citet{Nishiyama06a} derived
  $A_\J/A_\Ks \approx3.02$ and $A_\HH/A_\Ks\approx1.73$
  for the sightlines toward the GC based on red clump stars.
  They found a steep power-law for the near-IR
  extinction ($A_\lambda \propto \lambda^{-1.99}$)
  while in literature it is often thought that the near-IR
  extinction is a ``universal'' power-law
  $A_\lambda \propto \lambda^{-\beta}$
  with $\beta\approx 1.6-1.8$ \citep{Draine89, Draine03}.
  }

\section{Tracers: Red Giants and Red Clump Stars} \label{tracer}
\subsection{Red Giants}\label{Criteria_RG}
In the IR, red giants are appropriate tracers of interstellar
extinction\footnote{%
   Red giants were used as tracers to derive the extinction at 7$\mum$
   and 15$\mum$ by \citet{Jiang03,Jiang06}. The results were in close
   agreement with that from the hydrogen recombination lines
   \citep{Lutz96, Lutz99}.
   }
for the following reasons: (1) they have a narrow range of effective
temperatures so that the scatter of the intrinsic color indices is
small. The $\J-\Ks$ color index, often chosen to describe
$\lambda_{c}-\lambda_{r}$, has a scatter of only $\sim$\,0.1\,mag
around the central value of $\sim$\,1.2; (2) they are bright in the
IR and remain visible even with large extinction
and/or at a great distance.\footnote{%
  With $M_\Ks\sim-5.0$\,mag and $m_{\Ks}\sim13$\,mag,
          the distance can be as large as $d\approx 40$\,kpc,
          provided there is no extinction along the line of
          sight toward to the star.
         }
On average, their absolute $\Ks$ magnitudes
are as bright as $M_\Ks\sim-5.0$\,mag,
making them readily detectable by 2MASS even at a distance to the GC
($\simali$\,8.5$\kpc$). But we should note that evolved red giants
may have a circumstellar dust shell which would cause circumstellar
extinction and produce IR emission, affecting our understanding of
their intrinsic color indices. Accordingly, the selection of red
giants is usually based on the mid-IR colors which are barely
affected by interstellar extinction. Following
\citet{Jiang03,Jiang06}, we adopt the following criteria to select
red giants as our samples ---
\begin{enumerate}
\item $[3.6]-[4.5] <$ 0.6 and $[5.8]-[8.0] <$ 0.2,
    also adopted by \citet{Flaherty07},
    we confine ourselves to colors
    bluer than certain values to exclude the sources with IR
    excess such as pre-main sequence stars and asymptotic
    giant branch (AGB) stars. This criterion also
    safely excludes young stellar objects \citep{Allen04, Megeath04}.
    However, it is more complicated for evolved stars.
     By analyzing the synthetic colors of AGB stars obtained by
     convolving the \emph{ISO} Short Wavelength Spectrometer (ISOSWS)
     spectra with the IRAC transmission profiles,
    \cite{Marengo07} found that the IRAC colors of AGB stars
     are similar to that of red giant branch (RGB) stars.
     Therefore the selected samples may be contaminated
     by some AGB stars.
    \citet{Groenewegen06} also found $[3.6]-[4.5] > 0.3$
    for AGB stars with a significant mass loss.
    The $[5.8]-[8.0]$ color index is slightly more selective,
    with young AGB stars being redder by $\simali$0.2 than
    most red giants. Nevertheless, the criteria are kept with
    significant confidence as red giants are much
    more numerous than AGB stars in
    these relatively blue colors.
    This is later proved to be correct in the appearance
    of the color-color diagram of the sources.
    But it should be kept in mind that some
    AGB stars that suffer circumstellar extinction
    may be included.\footnote{%
       We don't know the exact fraction of stars
               in these colors that are contaminated by
               AGB stars. But considering a star of 1.5
               solar mass, it spends about $7.6\times 10^8$\,yrs
               at RGB and $\sim10^7$\,yrs at AGB \citep{VassWood1993}.
               The number ratio of RGB/AGB would be about
               two orders of magnitude.
               In addition, their color index $[5.8]-[8.0]$ of AGB stars
               ranges from about 0.0 to 1.0 \citep{Marengo08},
               mostly much redder than our critical value 0.2.
               Combining these two factors, it is a reasonable
               estimation that the fraction of AGB stars in the
               selected sample should be less than 1 percent.
               }
\item $\J-\Ks >$\,1.2 or $\HH-\Ks >$\,0.3.
    This criterion excludes foreground dwarf stars
    since even late-M dwarfs have $\J-\Ks <$\,0.6.
    Both theoretical and observational studies suggest that
    RGB stars have an intrinsic color of $\J-\Ks \approx 1.2\magni$
    with a dispersion of $\simali$0.1\,mag \citep{Glass99,Bertelli94}.
    For an M5 giant, the calculated $\J-\Ks$ is $\simali$1.2\,mag,
    and $\HH-\Ks$ is $\simali$0.3\,mag.\footnote{%
      If the $\HH$ band is chosen as the comparison band, this criterion
      ($\HH-\Ks >$\,0.3) is comparable to that of \citet{Flaherty07},
      i.e. $\HH-\Ks >$\,0.2.
      }
   {\it In the following, the samples selected in terms of
   $\J-\Ks >$\,1.2 are denoted by RG1,
   while the ones selected from $\HH-\Ks >$\,0.3
   are referred as RG2.}
\item Good photometric quality. This criterion guarantees the
      reliability of the calculated color indices without reducing
      the number of sources too much so as to
      degrade the statistical accuracy.
      Specifically, the photometry quality flags
      in the 2MASS $\J\HH\Ks$ bands are ``AAA''
      (i.e. with S/N\,$\geq$\,5);
      the quality flags in the IRAC bands are comparable
      (in the GLIMPSE catalogs the photometric uncertainty
       is typically $<$\,0.2\,mag).
      For all the 131 sky fields studied here,
              S/N\,$\geq$\,5 is required
              in all three 2MASS bands and four IRAC bands.

\item A deviation of $<3\sigma$ from the line
      fitted to the observed colors $\Ks - \lambda$ ($\lambda$ is the IRAC band wavelength)
      vs. $\J-\Ks$. 
      We used the IDL robust fitting program to
      iterate the fit, rejecting the sources
      with a deviation larger than 3$\sigma$.
      Those sources (with a deviation larger than 3$\sigma$)
      were rejected because, from a statistical point of view,
      it is unlikely for them to follow the linear relation.
      Furthermore, most of them have relatively large
      $\Ks - \lambda$ values, indicating the possible presence of extinction
      arising from circumstellar envelopes \citep{Jiang06}.
      \citet{Indebetouw05} also rejected high-$\sigma$ points
      to exclude some of the extreme red excess sources.
\end{enumerate}

Figure\,\ref{fig1} shows the near-IR color-magnitude diagrams (CMDs)
of two typical fields, $l=309^{\circ}$ to $310^{\circ}$ and
$l=11^{\circ}$ to $12^{\circ}$. In the left panels (a, c), the RG1
samples (based on $\J-\Ks >$\,1.2 and the other criteria described
above) are denoted by red dots. The black background points are the
sources in those fields with S/N\,$\geq$\,1 in all three 2MASS
bands. Because we only adopt the sources with S/N\,$\geq$\,5 in all
7 bands (3 2MASS bands and 4 IRAC bands), the number of the selected
RG samples is much smaller than the total number of sources in these
fields.

\subsection{Red Clump Giants}\label{Criteria_RCG}
\subsubsection{Red Clump Giants as a Tracer}\label{RCG}
\citet{Indebetouw05} chose red clump giants (RCGs), which have a
dispersion of $\simali$0.3\,mag in the absolute $\Ks$ magnitude and
$\simali$0.2\,mag in the near-IR color index \citep{Lopez02}, as a
tracer to derive the extinction from the 2MASS $\J$ band to the IRAC
8.0$\mum$ band. RCG stars are the equivalent of the
horizontal-branch stars for a metal-rich population. They have
narrow distributions in luminosity and color indices, with a weak
dependence on metallicity \citep{Nishiyama06a}. Thus their locations
in the near-IR CMDs could be more clearly distinguished than that of
red giants.
This makes them a more homogeneous sample of tracers to
obtain the IR extinction.\footnote{%
   RCG stars have long been used as a reliable standard candle to
           study Galactic structure (e.g. see \citealt{Demers97, Lopez02, Nishiyama06b}),
           as the reddening along the line of sight toward RCG
           stars is commonly considered to be almost
           completely due to interstellar extinction.
           Precisely speaking, from the view of the evolutionary
           stage of RCG stars, they should be at the post-RGB phase,
           and small amounts of dust should be there arising from
           the small mass loss at the RGB phase. But this would not affect our results since they have
           very similar color indices (this earns them the name
           ``clump''). Therefore, it is secure to attribute
           any excess color reddening to interstellar extinction.
   }

However, RCG stars are fainter than red giants in the IR because of
their relatively blue color index (usually K-type) and lower
absolute luminosity [$M_\Ks \sim -1.65$\,mag, see
\citet{Wainscoat92}; or $M_\Ks\sim-1.61\pm0.01$\,mag, see
\citet{Alves00}]. The RCG stars detected by 2MASS are not as distant
as red giants. Therefore, they can not trace the extinction as deep
as red giants. Due to the weakness of the IR extinction, the
traceable depth of extinction is crucial. As shown in
\citet{Indebetouw05}, the maximum $E(\J-\Ks)$ traced by RCG stars
was about 2.6\,mag, much smaller than the easily-reached 6\,mag of
red giants.  In addition, the color dispersion of RCG stars,
$\simali$0.2\,mag, is a little bit larger than that of red giants,
$\simali$0.1\,mag \citep{Glass99,vanLoon03}. Nevertheless, in this
work RCG stars are also selected as a tracer in deriving the IR
extinction and the results are compared with that from the red giant
samples.

As K2III giants, the commonly accepted absolute magnitude for RCG
stars in the $\Ks$ band is $M_\Ks \approx -1.65\magni$ and their
intrinsic IR color indices are almost constant, with
$(\J-\Ks)\approx0.75$ \citep{Wainscoat92}. These characteristics
would place them in a well defined narrow stripe in the near-IR CMDs
($\J-\Ks$ vs. $\Ks$). However, the distribution of RCG stars in the
CMD (based on their {\it observed} magnitudes and colors) is much
more scattered. This is because the observed color indices depend
only on interstellar extinction, while the observed $\Ks$ magnitudes
depend not only on the amount of extinction but also on the distance
of the star. In addition, the extinction is highly uneven and is not
simply proportional to the distance. Furthermore, there might be
contamination of dwarf stars and AGBs
at the lower end in the CMDs.\footnote{%
           \citet{Lopez02} found that for $\Ks < 12.5$
           only $\sim$\,2.5--5\% of the detected sources are dwarfs,
           but this fraction rises to $\sim$\,10--40\% for
           $13 < \Ks < 14$ \citep{Lopez02, Cabrera-Lavers07}.
   }
Thus the selection of the RCG stars appears empirical and eye-dependent.

\begin{figure}
 \centering
 \includegraphics[angle=0,width=5in]{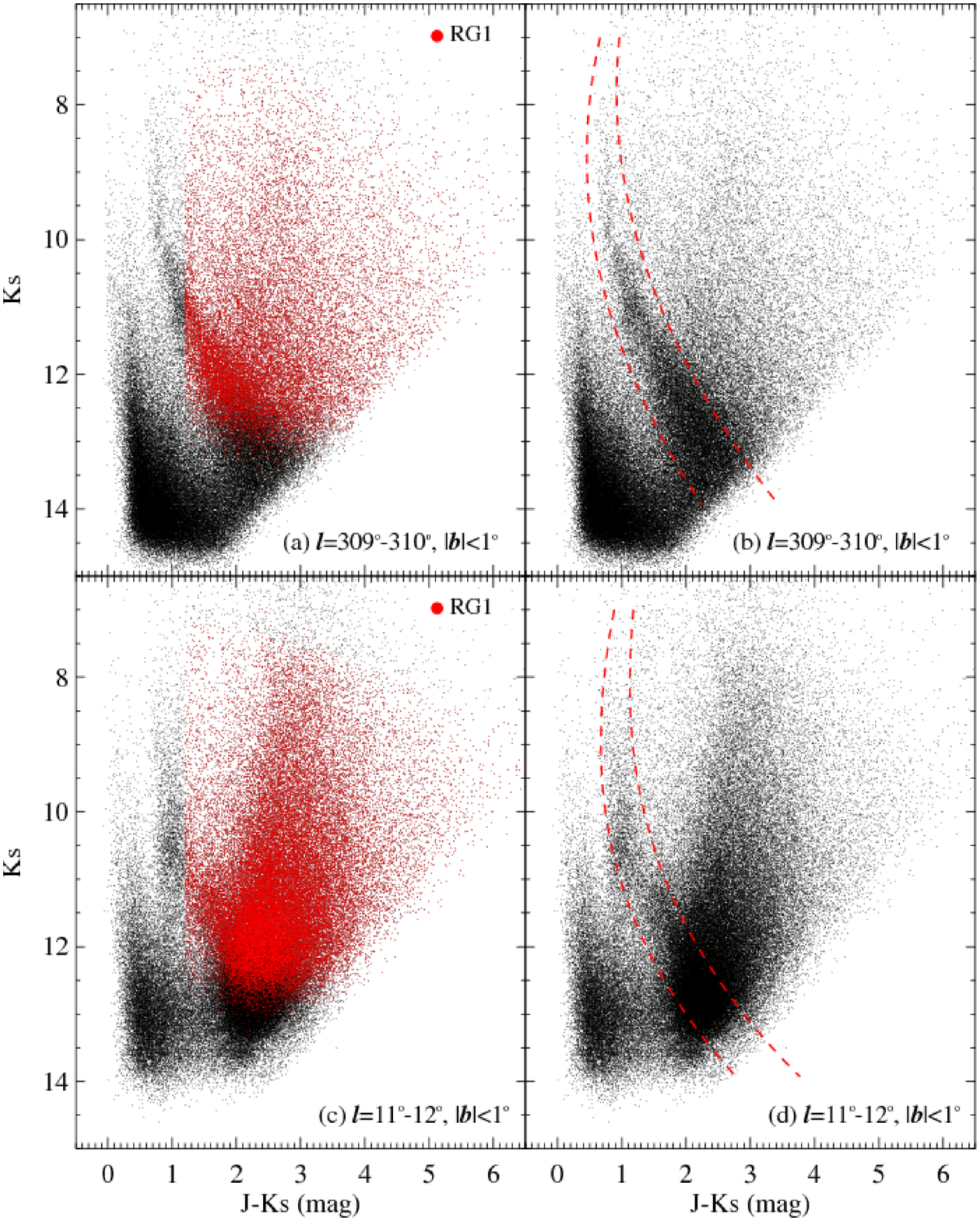}
 \caption{\footnotesize
          Color-magnitude diagrams for two GLIMPSE fields:
          $l=309^{\circ}$--$310^{\circ}$ and $|b|< 1^{\circ}$
          (upper panels: a, b), and $l=11^{\circ}$--$12^{\circ}$
          and $|b|<1^{\circ}$ (lower panels: c, d).
          Denoted by black dots, the sources in these fields
          with S/N\,$\geq$\,1 in all three 2MASS bands are all plotted.
          In the left panels (a,\,c), red dots denotes the red giants
          selected satisfying $\J-\Ks$\,$>$\,1.2 (RG1),
          S/N\,$\geq$\,5 and the other criteria
          (see \S\ref{Criteria_RG}).
          In the right panels (b,\,d), the selected RCG stars
          lie in between the two red dashed lines,
          with the center characterized by
          $(\J-\Ks)=5.28-1.11\Ks+0.07\Ks^2$
          and $(\J-\Ks)=7.09-1.44\Ks+0.08\Ks^2$ for
          $l=309^{\circ}$--$310^{\circ}$
          and $l=11^{\circ}$--$12^{\circ}$, respectively.
          The half widths of the RCG stripes are
          $\simali$0.3$\magni$ at $\Ks=13$ for panel b,
          $\simali$0.2$\magni$ at $\Ks=13$ for panel d.}
          \label{fig1}
\end{figure}

In Figure\,\ref{fig1}, the right panels show the RCG stripes lying
between two red dashed lines. The extension of the color index
$(\J-\Ks)$ of red giants (see the left panels of Fig.\,\ref{fig1})
clearly connects to that of red clump stars.
The upper right CMD (Fig.\,\ref{fig1}b)
has a relatively clear branch of red clump stars,
but this is not the case for the bottom one. In comparison
with the quite scattered distribution of red giants, the disk RCG
stars stand out by their configured locations in the CMDs for a
large portion of the sightlines (e.g. the field $l=309^{\circ}$ to
$310^{\circ}$), i.e. a narrow stripe from upper-left to lower-right
in the CMDs caused by interstellar extinction and the increasing
distance from the Sun. Within the RCG strip, the red clump stars consist of
relatively homogenous samples. However, the homogenity depends very
much on the specific field, (e.g. the field with $l=11^{\circ}$ to
$12^{\circ}$ and $|b|\leq 1^{\circ}$ does not present a clear stripe
of RCG stars in Figure\,\ref{fig1}d). The interstellar extinction is
possibly highly nonuniform with distance on the sightline within
$l=11^{\circ}$ to $12^{\circ}$, so that a gap appears in the RCG
stripe near $(\J-\Ks)\simeq1.5$ and $\Ks\simeq11$. In addition, a
giant branch can be seen at $\J-\Ks\approx2.8$ in the lower CMD in
Figure\,\ref{fig1}d, which is probably due to the stars in the bulge
as suggested by \citet{Hammersley00}.

Although the 2MASS Point Source Catalogue is complete down to $\Ks
\approx 14.3\magni$ \citep{Cutri03}, the faintest $\Ks$ magnitude in
this work is $\simali$13$\magni$ for almost the entire GLIMPSE
fields (because of the S/N\,$\geq$\,5 requirement for all the 2MASS
and IRAC bands). The faintest $\Ks$ magnitude of the selected RG
samples can only reach $\simali$13$\magni$ (see Fig.\,\ref{fig1}a,
c). For the RCG stars selected from the RCG stripes requiring
S/N\,$\geq$\,5 in all seven bands, the faintest $\Ks$ magnitude can
also only reach $\simali$13$\magni$.

\subsubsection{Selection of RCG Stars}\label{method_RCG}
\citet{Lopez02} proposed a method to extract RCG stars based on
CMDs. They determined an empirical track of RCG stars by linking the
peaks of the histograms of all the horizontal cuts in the CMDs.
Applying a limitation of $\Ks<$\,13.0 to exclude K dwarf stars, the
sources within a deviation of $<$\,0.2 mag in $\J-\Ks$ from the
determined track are extracted as the selected RCG stars.
\citet{Drimmel03} and \citet{Indebetouw05} adopted similar methods
to pick up RCG stars. In this work, we take a similar approach.

First, a rough stripe in the $\J-\Ks$ vs. $\Ks$ CMD was chosen by
eye to encircle the preliminary range of the RCG stars. The
eye-selected range is divided into different horizontal cuts with a
step of 0.3\,mag at $\Ks$, then the histograms of each horizontal
cut was fitted with Gaussian functions to determine
the peak color indices $\J-\Ks$ for different horizontal $\Ks$ cuts.
Second, the peak positions in each horizontal cuts are taken as
input to delineate the curve fitted with a 2nd-order
polynomial for the central location of the RCG stars.\footnote{%
   Please contact the authors for the parameters
   if one is interested in carrying out similar
   computations.
   }
Third, the width of the curve derived from the previous step, which
corresponds to the scale of the dispersion of the color index, is
determined. Unlike \citet{Lopez02}, we do not treat this width as a
constant (because it results not only from the almost constant
scatter of the intrinsic color index, but also from the photometric
error which increases with decreasing apparent brightness, i.e.
increasing magnitude). The increase of dispersion width with the
observed magnitude is noticeable from the CMDs as well. So the width
adopted in our work differs for different sightlines based on the
definition of the RCG stripes in the CMDs. Centered with the fitted
red clump track, the width ranges from $\simali$0.1 to
$\simali$0.2\,mag for $\Ks=8.0$ and increases to
$\simali$0.2--0.3\,mag for $\Ks=13.0$ (see Fig.\,\ref{fig1}).

Finally, one more factor to take into account is the enhanced
contamination of dwarf stars at fainter magnitudes. These dwarfs
should be mainly earlier than K-type to be bright enough to be
visible at relatively large distances (because their observed colors
are much redder than the intrinsic ones which must result from
subjecting to substantial extinction, indicating that they can not
be nearby). The near-IR intrinsic colors of these dwarf stars are
bluer than that of the RCG stars. Compared with the RCG stars in the
mid-IR, these dwarfs would have a larger $k_{x}$ in equation
(\ref{slope}) and lead to a smaller $A_{x}/A_{r}$ in equation
(\ref{ext}) since $A_{c}/A_{r}>1$. To suppress this effect, the
$\Ks$ magnitude is limited to 13.0 in \citet{Lopez02}. We also place
an upper limit on the $\Ks$ magnitude in order to
reduce the confusion from the K dwarfs\footnote{%
          To suppress the contaminations of YSOs and AGB stars,
          we adopt the same criterion on the IRAC color index
          as we did to select the RG samples,
          i.e. $[3.6]-[4.5] <$\,0.6 and $[5.8]-[8.0] <$\,0.2
          (see \S\ref{Criteria_RG}).
   }
(but this upper limit is again different for different sightlines).
As mentioned earlier, the $\Ks$ magnitude can only reach
$\simali$13$\magni$ for most of the GLIMPSE fields because of the
S/N\,$\geq$\,5 requirement for all seven bands (see \S\,\ref{RCG}).
Consequently, for most of the sky fields, the typical limit on the
$\Ks$ magnitude is $\simali$12\,mag. But this limit is much smaller
on the sightlines toward the Galactic bulge direction, mostly in the
range of $|l|<15^{\circ}$, where $\Ks \leq$ 11.5 is generally taken.
The lower cut of the $\Ks$ magnitude results in much smaller
$\J-\Ks$ and much fewer samples than that from the red giant
samples. Furthermore, because of the lower $\Ks$ magnitude
($\Ks$\,$\leq$\,12), the RCG samples in this work can only be used
to probe the interstellar extinction from
the diffuse ISM near the Sun.\footnote{%
   The RCG stars in the Galactic bulge
           ($|l|<20^{\circ}$) appear to lie at
           a larger distance \citep{Nishiyama06a}
           and have a $\Ks$ magnitude of
           $\simali$12--14 \citep{Hammersley00}.
           Because of the $\Ks$ magnitude limitation,
           the RCG stars finally extracted in this work
           would mostly lie in the Galactic disk.
   }
According to the following expression,
\begin{equation}\label{distance}
5\,\log d = m_\Ks - M_\Ks + 5 - A_\Ks(d),
\end{equation}
if $A_\Ks\approx1$, the limitation of $\Ks \leq12$ implies that
the most distant stars that could be selected in our samples are
at $d\sim 3.5$\,kpc.
Many works have suggested the existence of
a linear bar across the Galactic center with
a half-length of $\simali$4.5\,kpc and a position angle of
$\simali$45$^{\circ}$
\citep{Hammersley94,Hammersley00,Benjamin05,Cabrera-Lavers07},
i.e. the closest distance to the sun is $\sim5.3\kpc$ (assuming the distance to the GC from the sun is $R_{\rm sun}\approx 8.5\kpc$).
Thus even if $A_\Ks\approx0$, we can still hardly reach the bulge
stars located in the bar beyond $d\sim5$\,kpc.
In the region within $|l|<15^{\circ}$, where $\Ks
\leq 11.5$ is taken, the distance limit must be closer than
$\sim$\,4\,kpc. So the interstellar extinction derived from the RCG
samples is mainly from the local diffuse ISM and it may differ from
that derived from red giants (see \S\ref{variation-l}).

For most of the 131 GLIMPSE fields, the stripes of the RCG stars in
the near-IR CMDs are clearly visible, indicating that the
above-described selection method for RCG stars is
reasonably reliable. But for
the fields in the directions to the Galactic bulge, there are so
many dwarf stars that the stripes of RCG stars are not easily
distinguishable, thus only the nearby RCG stars subject to small
extinction which lie at the higher end of the stripes can be
extracted. Moreover, in the regions within $|l|< 15^{\circ}$, there
is always a giant branch overlapping with the RCG stripe at the low
$\Ks$ end (see the vertical branch near $(\J-\Ks)=2.8$ in
Figure\,\ref{fig1}d). Significant contamination occurs at
later-type RGB and AGB stars \citep{Hammersley00, Indebetouw05}. The
contamination caused by other types of stars degrades the quality of
the results derived from the RCG samples. The reliability of the
results varies from field to field.

\section{Results and Discussion}\label{discussion} \label{results}
\subsection{Mean Extinction} \label{average}
\begin{figure}
 \centering
 \epsscale{0.7} \plotone{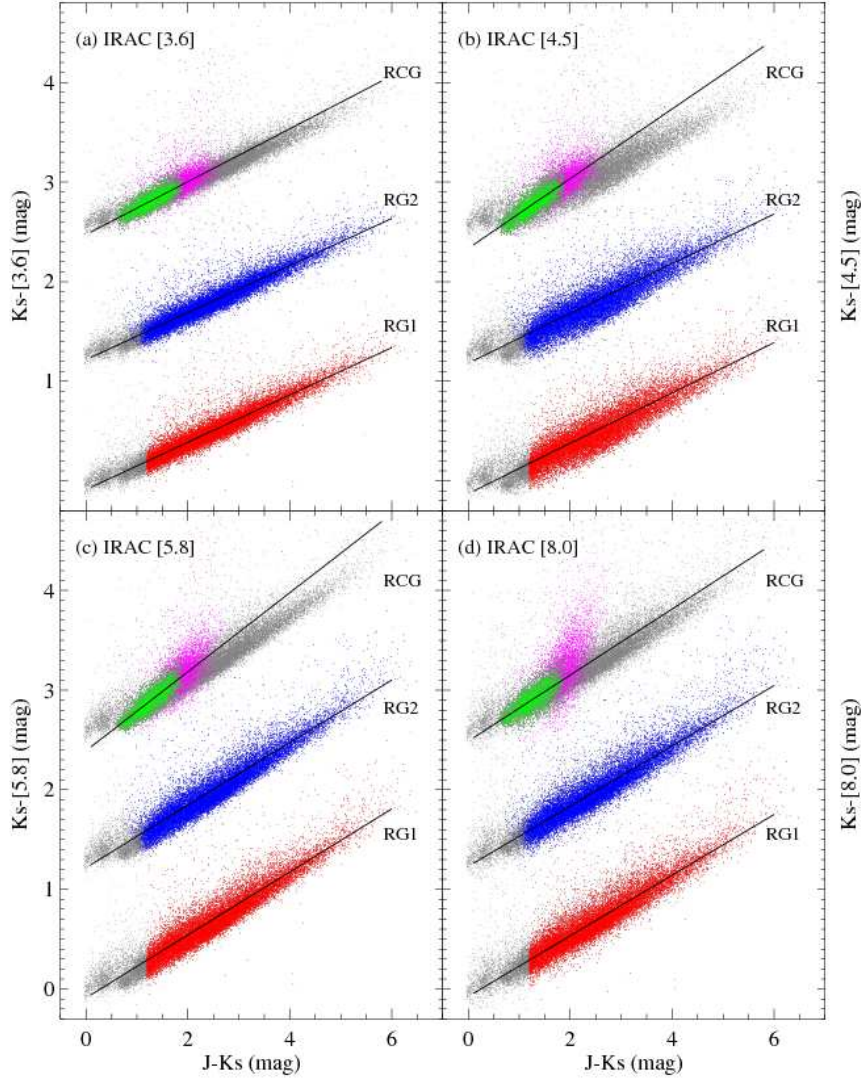}
 \caption{\footnotesize
          Color-color diagrams for the red giants (RG1, RG2) and
          red clump giants (RCG) selected in a typical sightline,
          with $l=309^{\circ}$--$310^{\circ}$ and $|b|< 1^{\circ}$.
          For clarity, the RG2 and RGC samples are vertically
          shifted +1.3$\magni$ and +2.6$\magni$, respectively.
          The black line is a linear fit to the color indices.
          For the RG1 and RG2 samples, the two fitted black lines
          are parallel to each other
          (this is because the RG1 and RG2 samples
           are almost identical and they are computed using
           the same comparison band $\J$).
          The RCG samples are rarer and
          have a smaller range of $(\J-\Ks)$.
          The magenta dots plot
          all the sources with S/N\,$\geq$\,5
          in the RCG stripe in Figure\,\ref{fig1}b,
          while the green ones are further limited to $\Ks \leq 12$.
          In panel d,  the fitted line derived
          from the RCG samples could be distorted by
          the contamination of YSOs and AGB stars.
          This is why we limit the $\Ks$ magnitude to 12
          for this GLIMPSE field. Gray dots plot all sources detected by GLIMPSE with S/N\,$\geq$\,5 in all bands.
          \label{fig2}}
\end{figure}

We have obtained the extinction in each IRAC band
(relative to that of the $\Ks$ band)
averaged over all 131 GLIMPSE fields
(see Table\,\ref{tbl-relext}).
With red giants as the extinction tracer
and the $\J$ band as the comparison band,
the mean extinction ratios
are $A_{[3.6]}/A_{ \Ks}\approx0.64\pm0.01$,
$A_{[4.5]}/A_{\Ks}\approx0.63\pm0.01$,
$A_{[5.8]}/A_{\Ks}\approx0.52\pm0.01$ and
$A_{[8.0]}/A_{\Ks}\approx0.55\pm0.02$. The selection criteria of
$\J-\Ks >1.2$ (RG1) or $\HH-\Ks>0.3$ (RG2) makes
little difference since the samples selected from these two criteria
are almost the same (see Fig.\,\ref{fig2} and the first two rows of
Table\,\ref{tbl-relext}). Here we adopt $A_\HH/A_\Ks = 1.56$ and
$A_\J/A_\Ks =2.52$ (i.e. we take the power index $\beta$ of the
near-IR extinction law $A_\lambda \propto \lambda^{-\beta}$ to be
1.61; see \citealt{Rieke85,Draine03}).
With the $\HH$ band chosen as the comparison band, the extinction
ratios $A_\lambda/A_\Ks$ (derived from the RG2 samples, see the
third row of Table\,\ref{tbl-relext}) systematically decrease by an
amount of $\simali$0.04 at 3.6$\mum$, $\simali$0.04 at $4.5\mum$,
$\simali$0.05 at $5.8\mum$, and $\simali$0.05 at 8.0$\mum$,
respectively.
If we take a larger power index
[e.g. $\beta = 1.75$ as suggested by \citet{Draine89}],
the extinction ratios will become smaller.\footnote{%
 This is true no matter which band, $\J$ or $\HH$,
         is chosen as the comparison band.
         But the effect is more significant
         for the former and therefore there will be
         little difference between the extinction ratios
         derived using different comparison bands ($\J$ or $\HH$).
         }

Also shown in Table\,\ref{tbl-relext} are the extinction ratios
$A_\lambda/A_\Ks$ derived from the RCG samples (with the $\J$ band
as the comparison band). The extinction ratios of the 3.6$\mum$,
4.5$\mum$ and 5.8$\mum$ bands derived based on the RCG samples are
smaller than that based on red giants, while the extinction ratio of
the 8$\mum$ band is approximately equal for both
samples.\footnote{%
  We should note that the approximate equality between the extinction from
  the red giant and RCG samples in the 8.0$\mum$ band is an average result.
  As far as individual regions are concerned, the difference in this band
  is significant in the Galactic bulge region (see \S\ref{variation}).
  }
As mentioned earlier, the RCG samples may include some
dwarf stars at the faint end. Since dwarf stars are intrinsically
bluer than RCG stars, they may cause the derived extinction
to be underestimated (see eq.\ref{ext}). On the other hand, the
extinction would be somewhat overestimated from the red giant
samples because of their circumstellar dust. But for the
8.0$\mum$ band, the extinction derived from red giants may be
underestimated because of the possible presence of silicate emission.
\footnote{%
  The red giants (mostly M-type stars) selected here
  may have a very thin shell of silicate dust
  and therefore it is likely that they exhibit a weak 9.7$\mum$
  silicate emission feature. The uncertainties discussed here
  are not included in the quoted errors in Table 1.
  }

\begin{figure}
 \centering
 \includegraphics[angle=0,width=3.5in]{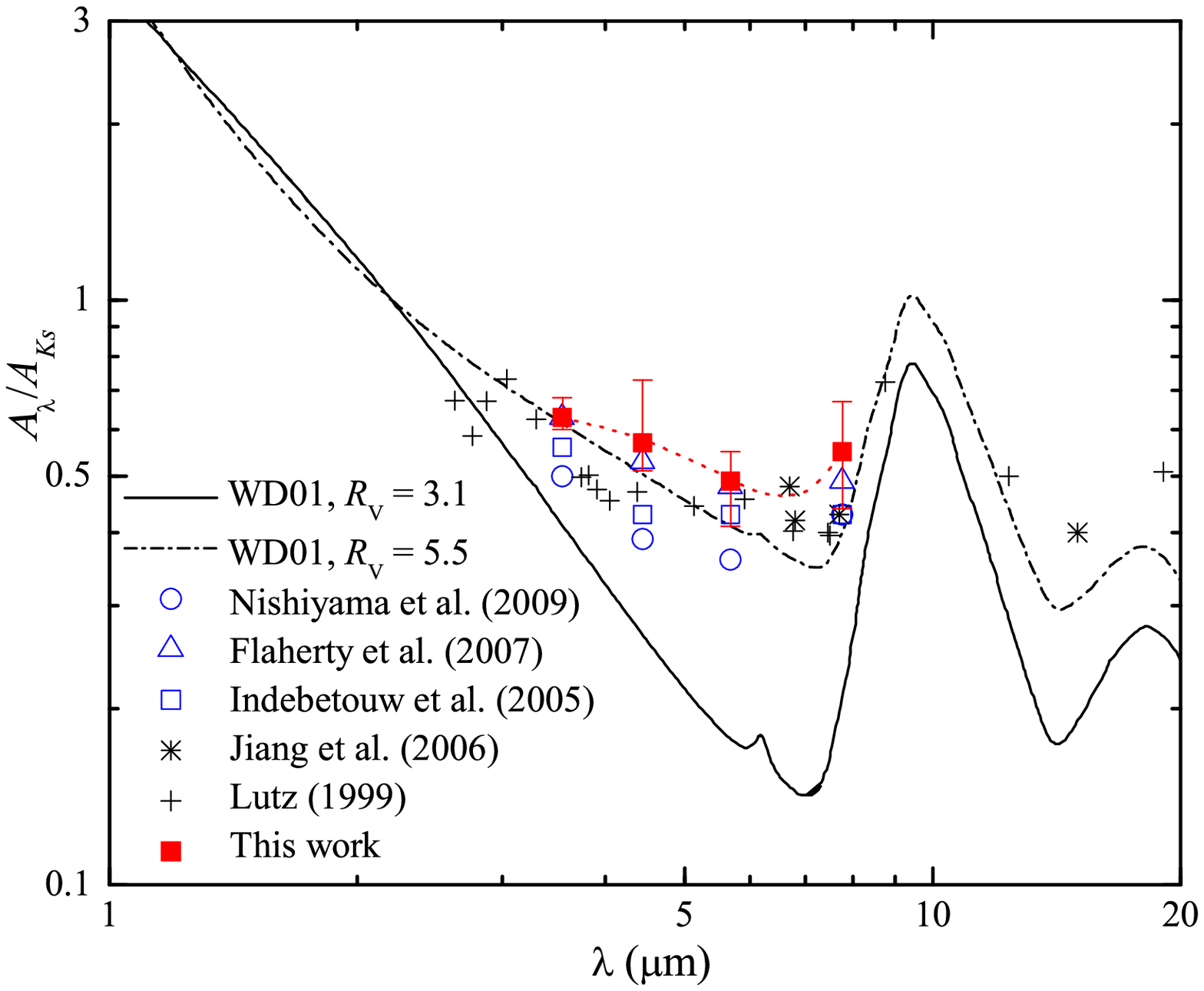}
\caption{\footnotesize
         Comparison with previous observational determinations based on
         {\it ISO} and {\it Spitzer} data and with the extinction curves
         calculated from the interstellar grain model for $R_V=3.1$
         (solid line) and for $R_V=5.5$ (dot-dashed line) of
         \citet{Weingartner01} (WD01). The dotted line smoothly connects
         the mean extinction ratios $A_\lambda/A_\Ks$
         derived in this work and that of \citet{Jiang06} at 6.7$\mum$.
         The red bars plot the uncertainty ranges of
         $A_\lambda/A_\Ks$ in each IRAC band.
\label{fig3}}
\end{figure}

Both \citet{Indebetouw05} and \citet{Flaherty07} have studied the
extinction in the GLIMPSE OSV field. To allow a direct comparison
with their results, in Table\,\ref{ext_OSV} we tabulate the IRAC
band extinction (relative to $A_{\Ks}$) derived here as well as
theirs. The extinction values taken from \citet{Indebetouw05} are
for the off-cloud line of sight and derived from RCG stars. Although
the extinction results are similar in the IRAC [3.6] and [4.5]
bands, the extinction law derived here from the RCG samples for the
OSV field is much steeper than that derived by \citet{Indebetouw05},
with much smaller values in the [5.8] band. This is probably due to
the different methods adopted in selecting RCG stars, such as the
limitation on the $\Ks$ magnitude or the criterion on color index
(see \S\,\ref{method_RCG}). With the same criteria ($[3.6]-[4.5]
<$\,0.6 and $[5.8]-[8.0] <$\,0.2, RG2) and the same comparison band
$\HH$ as adopted by \citet{Flaherty07}, our results are $\sim$0.3
larger in the [4.5], [5.8] and [8.0] bands (see the third row of
Table\,\ref{ext_OSV}). The reason for this systematic
difference could be attributed to the different power index $\beta$
used as discussed above. Nevertheless, the major reason why the
extinction derived here (the first and second rows in
Table\,\ref{ext_OSV}) is systematically larger than that derived by
\citet{Flaherty07} for the OSV field lies in the choice of the
comparison band: in this work we use the $\J$ band as the comparison
band, while \citet{Flaherty07} used the $\HH$ band.
In view of the different characteristics of RG and RCG stars, we
averaged over both the RG1 and RCG results in order to get a
``mean'' extinction law more or less unrelated to the distance. The
``mean'' interstellar extinction in the four IRAC bands are
$\simali0.63\pm0.01$, $0.57\pm0.03$, $0.49\pm0.03$ and $0.55\pm0.03$
for $A_{[3.6]}/A_{\Ks}$, $A_{[4.5]}/A_{\Ks}$, $A_{[5.8]}/A_{\Ks}$
and $A_{[8.0]}/A_{\Ks}$, respectively (see Table\,\ref{tbl-relext}).
In Figure\,\ref{fig3} we plot our results together with previous
observational determinations
\citep{Lutz99,Indebetouw05,Jiang06,Flaherty07,Nishiyama09} and the
model extinction calculated for $R_V=3.1$ and $R_V=5.5$ by
\citet{Weingartner01}. It is seen in Figure\,\ref{fig3} that our
results are in close agreement with previous studies and confirm
that the extinction law at $\simali$3--8$\mum$ is almost
flat and lacks the minimum around 7$\mum$ predicted from
the silicate-graphite interstellar grain model for $R_V=3.1$,
but is close to (although systematically slightly higher than)
the $R_V=5.5$ model curve.

\subsection{Variation of IR Extinction} \label{variation}
\subsubsection{Variation and Amplitude} \label{variation-a}
The interstellar extinction law is long known to vary along
different sightlines in the optical and UV wavelength range, while
the extinction in the near IR was often thought to be ``universal''
(but this was recently questioned by \citealt{Nishiyama06a}). From
the ISOGAL surveyed fields, \citet{Jiang06} found marginal variation
in $A_{7\mum}/A_\Ks$. Due to the relatively large uncertainty,
the regional variation was not clear.

\begin{figure}
\centering
\includegraphics[angle=90,width=5in]{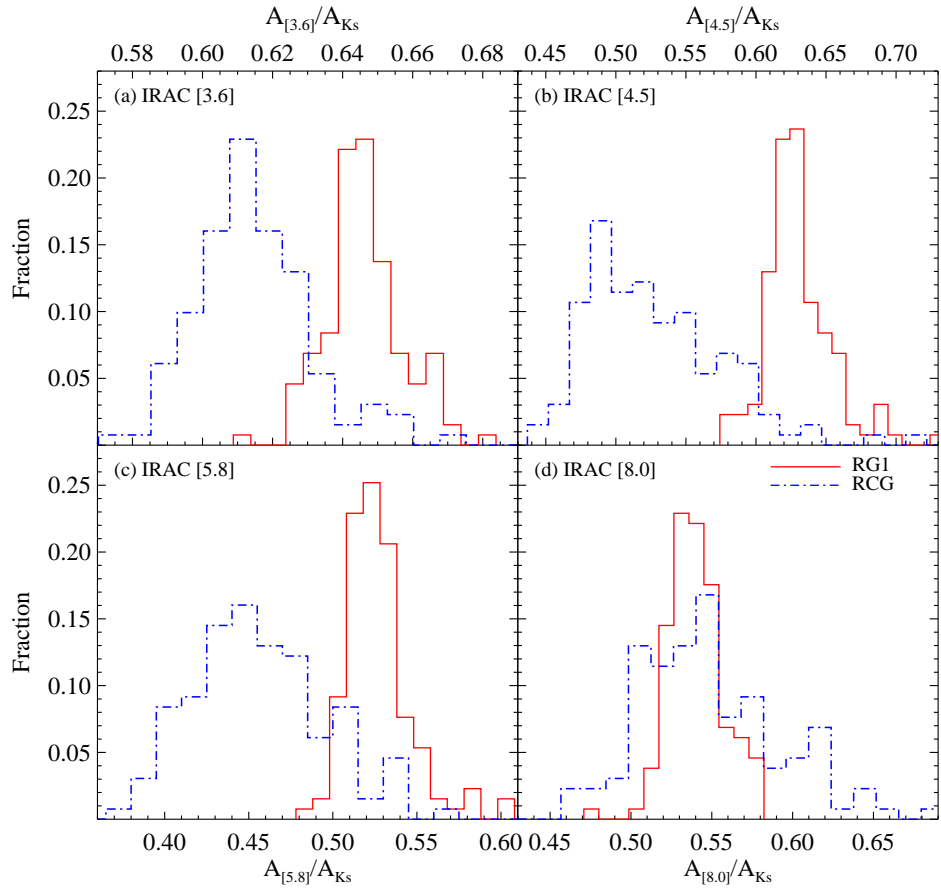}
\caption{\footnotesize
         Histograms of the extinction (relative to $A_\Ks$)
         of all 131 GLIMPSE fields in the four IRAC bands.
         \label{fig4}}
\end{figure}

To examine whether there exist regional variations of the wavelength
dependence of the IR extinction law, we show in Table\,\ref{131tab}
the extinction ratios $A_\lambda/A_\Ks$ in all four IRAC bands for
all 131 GLIMPSE fields, including results from both the RG1 and RCG
samples. Because of more sample stars, much larger $E(\J-\Ks)$ and
no limitation on the $\Ks$ magnitude, the typical error for the RG1
results is just $\simali$0.001, much smaller than that of the RCG
results which is $\sim$0.01. Figure\,\ref{fig4} plots the histograms
of the extinction ratios in all 131 GLIMPSE fields and clearly shows
that there are a range of extinction ratios for both the RG1 and RCG
results. The range for the RCG results is clearly larger than that
for the RG1 results. Even for the extinction ratios derived from red
giants, the range of variations for any of the IRAC bands are large
enough to exceed the average error of $\simali$0.001, with
$\simali$0.07, 0.16, 0.14 and 0.12 (see Table\,\ref{tbl-extrange})
at 3.6$\mum$, 4.5$\mum$, 5.8$\mum$ and 8.0$\mum$, respectively. This
demonstrates that the extinction (relative to $A_\Ks$) in the IRAC
bands varies among different sightlines. A similar conclusion has
already been drawn by \citet{Chapman09}, \citet{McClure09},
\citet{Nishiyama09}, and \citet{Fitzpatrick09}.

The amplitude of variation of $A_\lambda/A_\Ks$ is relatively small
for the 3.6$\mum$ band and is thus difficult to detect without high
quality data. This may explain why the extinction law in the near-IR
was thought to be ``universal'' \citep{Draine03}. The variations in
the 4.5$\mum$ and 5.8$\mum$ bands should be attributed to continuum
extinction, while the variation of the 8$\mum$ extinction largely
comes from the variation of the 9.7$\mum$ silicate absorption
feature strength (see Fig.\,\ref{fig3}) among different sightlines.
The fact that the 8$\mum$ dispersion is essentially no greater than
that of the 4.5$\mum$ and 5.8$\mum$ bands suggests that the
9.7$\mum$ silicate absorption feature probably does not vary much on
these sightlines.

\begin{figure}
\includegraphics[angle=0,width=3.5in]{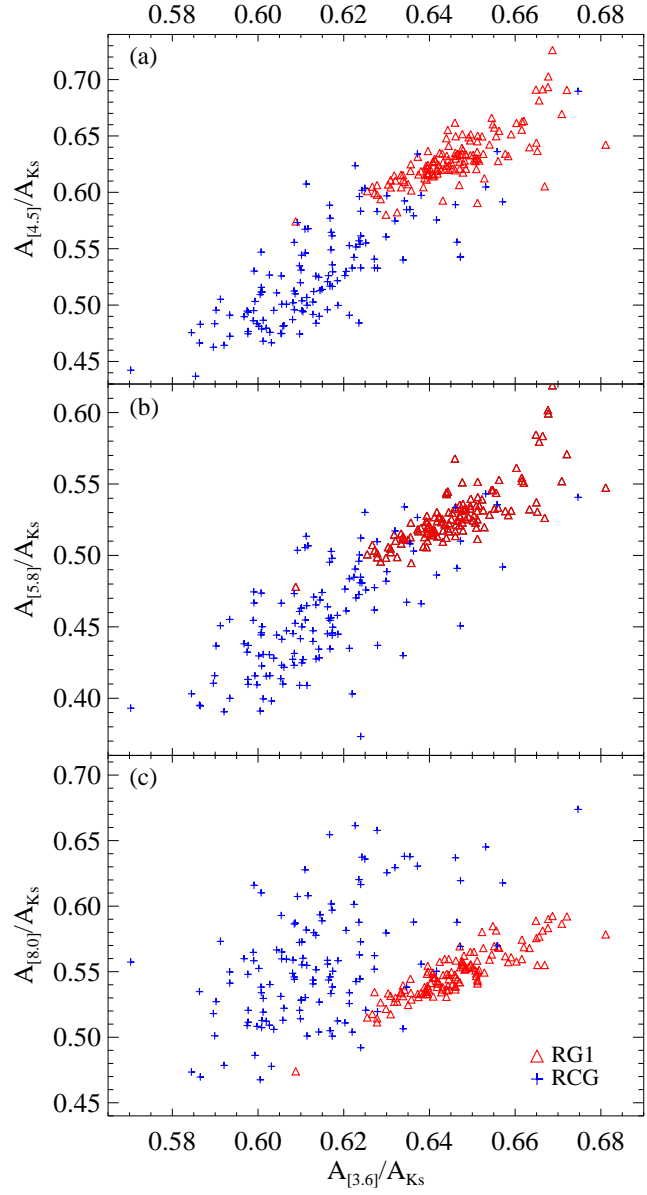}
\centering \caption{\footnotesize
         Correlation of $A_{[4.5]}/A_\Ks$, $A_{[5.8]}/A_\Ks$,
         and $A_{[8.0]}/A_\Ks$ with $A_{[3.6]}/A_\Ks$.
         Red, unfilled triangles are for RG1;
              blue crosses are for RCG.
         \label{fig5}}
\end{figure}

As shown in Figure\,\ref{fig5}, for the RG1 samples,
$A_{[4.5]}/A_{\Ks}$, $A_{[5.8]}/A_{\Ks}$ and $A_{[8.0]}/A_{\Ks}$ are
all closely correlated with $A_{[3.6]}/A_{\Ks}$, while the
correlation between $A_{[8.0]}/A_{\Ks}$ and $A_{[3.6]}/A_{\Ks}$ for
the RCG samples is not as tight as that of the [4.5] and [5.8] bands
with the [3.6] band. This is probably due to the contamination of
O-rich AGB stars at the lower end of RCG stripes: the extinction in
the 3.6$\mum$, 4.5$\mum$, and 5.8$\mum$ bands largely comes from the
dust which produces the $\simali$1--6$\mum$ continuum extinction
(e.g. graphite, see Fig.\,8 of \citealt{Draine84}), in contrast, the
8.0$\mum$ extinction arises mainly from the 9.7$\mum$ silicate
absorption feature. If the 3.6$\mum$, 4.5$\mum$, and 5.8$\mum$
extinction is caused by interstellar dust while the 8$\mum$
extinction is partly from interstellar dust and partly from O-rich
AGB stars, one would expect that $A_{[8.0]}/A_{\Ks}$ does not
correlate with $A_{[3.6]}/A_{\Ks}$ as closely as $A_{[4.5]}/A_{\Ks}$
and $A_{[5.8]}/A_{\Ks}$ correlate with $A_{[3.6]}/A_{\Ks}$.
We note that the absorption features of polycyclic aromatic
hydrocarbon (PAH) at 6.2, 7.7, and 8.6$\mum$ are too weak to
contribute to the discontinuities between RCG and RGs
(see Fig.\,16 and \S11 of \citet{Li01},
and \S10.3 of \citet{Draine07}.

\subsubsection{Variation with Galactic Longitude} \label{variation-l}
\begin{figure}
\hspace{-0.55in} \centering
\includegraphics[angle=90,width=5in]{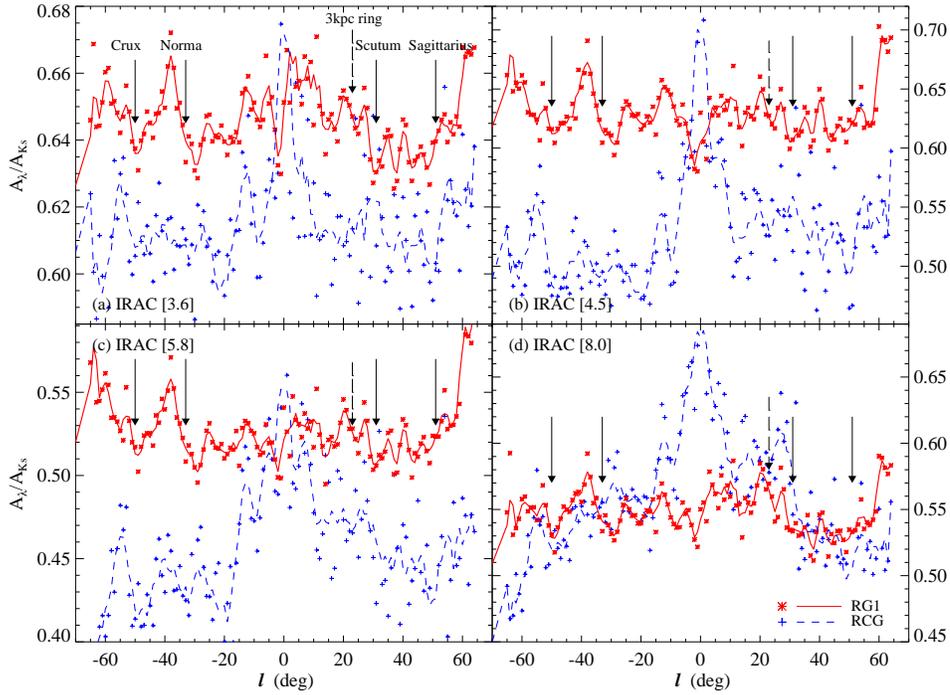}
\caption{\footnotesize
         Longitudinal distributions of $A_{\lambda}/A_\Ks$ in
         the four IRAC bands obtained from the RG1 (asterisks)
         and RCG samples (crosses).
         The typical error for the RG1 results is
         $\simali$0.001, which is much smaller
         than $\simali$0.01 for the RCG results.
         The solid and dashed lines are drawn with
         the smoothed results. The solid vertical arrows
         show the tangent directions of the spiral arms at
         $l=-50^{\circ}, -33^{\circ}, 31^{\circ}$
         and $51^{\circ}$ \citep{Vallee08}.
         The dashed arrow shows the tangent direction of
         the 3$\kpc$ ring at $l=23^{\circ}$ \citep{Dame08}.
         \label{fig6}}
\end{figure}

Unlike previous studies of the ISOGAL fields in which no clear
structural distribution along Galactic longitude or latitude was
found for the extinction at 7$\mum$ and 15$\mum$ \citep{Jiang06}, in
this work we see an interesting distribution of extinction.
Whichever extinction tracer [red giants (RG1, RG2) or RCG stars] or
sample-selection-criterion is used, we always see an uneven
distribution of extinction ratios in each IRAC band with
Galactic longitude (see Fig.\,\ref{fig6}). Since the results from
the RG1 and RG2 samples are approximately the same, we only show
that of the RG1 (and RCG) samples in Figure\,\ref{fig6}.

\paragraph{Longitudinal extinction profiles
           derived from RG1 or RCG samples:}
%
Although the extinction values derived from the RCG stars are
systematically smaller than that from the red giants as discussed
in the previous section,
the structural features appear at almost the same
positions outside the Galactic bulge region for both samples. The
common features include, dips around $l=-50^{\circ}$,
$l=-20^{\circ}$, $l=16^{\circ}$, $l=51^{\circ}$, peaks around
$l=-53^{\circ}$ (see Fig.~\ref{fig6}).
However, in the bulge direction within $|l|<15^{\circ}$,
the two samples yield different variation patterns.
This can be understood in view of the fact that
the RCG samples selected in this work do not trace as deeply as the
red giant samples to the Galactic bulge due to the limiting
magnitude of $\Ks<11.5$ on the RCG stars in these regions. As
mentioned earlier, the selected RCG samples can only reach a
distance of $\simali$4\,kpc from us to the bulge direction even if
there is no extinction.
In contrast, the red giants trace the structure of
the bulge such as the bar and the distant spiral arms,
which shows some of the differences
in the longitudinal distribution of $A_\lambda/A_\Ks$ toward the GC.
Another difference between the results from the RG1 and RCG samples
lies in the regions between $20^{\circ}<l<30^{\circ}$. Similarly,
the RCG samples can only probe the interstellar extinction of the
ISM of nearby spiral arms (e.g. the Sagittarius-Carina arm located
at $\simali$2\,kpc from the Sun), but not the extinction from
the more distant $\simali$3\,kpc-molecular ring between
$20^{\circ}<l<30^{\circ}$ which contains a significant
amount of CO and dust
\citep{Hammersley94,Dame01,Dame08}.\footnote{%
   \cite{vanWoerden57} suggested the apparent
           southern tangent for the near 3$\kpc$ ring
           is at $l=-22^{\circ}$.
           \citet{Cohen80} suggested the northern tangent
           is near $l=24^{\circ}$.
           \citet{Dame08} adopted
           tangent directions of $\pm23^{\circ}$
           for the 3$\kpc$ arm based on
           their composite CO survey.
           With $R_{\odot}=8.5\kpc$, the near arm is
           at a distance of $\simali$5.2$\kpc$ \citep{Dame08}.
    }
Meanwhile, the RCG stars can only trace the Crux-Scutum arm
at $l\approx-29^{\circ}$ \citep{Hammersley94}, not the more
distant Norma arm \citep{Vallee08}, thus the longitudinal
profiles from RCGs and RGs are also different
along this direction.

\paragraph{Relationship with the Galactic spiral arms:}
%
Because the dust properties of the spiral arm regions differ from
that of the inter-arm regions \citep{Greenberg95}, the variation of
extinction with Galactic longitude could be related to the spiral
structure.
We label the tangent positions of the spiral arms in
Figure\,\ref{fig6} with long black arrows.\footnote{%
  The number of spiral arms and other arm parameters
  (e.g. pitch angles) have not been well-determined, although
  there have been various investigations to determine them.
  For example, the number of spiral arms (2 or 4) is actively
  being debated (e.g. see \citealt{Benjamin08}). We favour
  the four-arm model for the sake of studying the dust extinction
  profile vs. the Galactic structure since the 240$\mum$ dust
  emission supports the four-arm model \citep{Drimmel00}. Note
  that both the extinction and 240$\mum$ emission are from dust.
  \citet{Vallee08} evaluated, compiled and compared previous works
  on the determination of the position parameters of the spiral arms.
  In his four-arm structure model, \citet{Vallee08} placed the mean
  longitudes of tangent from the Sun to the
  Carina-Sagittarius arm, the Crux-Scutum arm, the Norma arm, the
  start of the Perseus arm, the Scutum-Crux arm, and the
  Sagittarius-Carina arm at $l=-76^{\circ} \pm 2^{\circ}, -50^{\circ}
  \pm 3^{\circ}, -33^{\circ} \pm 3^{\circ}, -21^{\circ} \pm 2^{\circ},
  31^{\circ} \pm 3^{\circ}$ and $51^{\circ} \pm 4^{\circ}$,
  respectively (see his Table\,2 and Fig.\,2).
  }
It is seen that the locations of the spiral arms coincide with the
dips of the extinction ratios $A_\lambda/A_\Ks$ derived from the RG1
samples.\footnote{%
   Since the RCG samples can not probe the extinction to distant arms
   (therefore they can not reflect the large-scale structure of the
   spiral arms), we will only consider $A_\lambda/A_\Ks$ from the RG1
   samples in the following discussion.} In particular, the coincidence of the locations of the
$A_\lambda/A_\Ks$ minimum and the spiral arms is outstanding at
negative longitudes (e.g. the Crux-Scutum arm at $l=-50^{\circ}$,
the Norma arm at $-33^{\circ}$, and the southern tangent of the
3$\kpc$ ring around $-23^{\circ}$). For the positive longitudes, the
locations of the Scutum-Crux arm and the Sagittarius-Carina arm also
coincide with the valleys of $A_\lambda/A_\Ks$. For the
broad dip around $l=16^{\circ}$, there is no clear tangent direction
to any arms, but this direction points to the start of the Norma arm
and the Scutum-Crux arm (see Fig.\,2 of \citealt{Vallee08}),
and probably one end of the Galactic bar.\footnote{%
  In Vallee's model, the Galactic bar is simply drawn on
  his cartographic model with the mean values from
  the literature. \citet{Benjamin05} determined the radius
  ($R_{\rm bar} = 4.4\pm0.5\kpc$) and orientation
  ($\phi=44^{\circ}\pm10^{\circ}$) of the Galactic bar
  from the M and K giants in the GLIMPSE survey.
  }
\citet{Hammersley94} found an extra
dip around $l=16^{\circ}$ in the longitude distribution of the
2.2$\mum$ flux obtained by the Diffuse Infrared Background
Experiment (DIRBE) on board the \textit{Cosmic Background
Explorer} (\textit{COBE}) (see their Fig.\,1). They
speculated that it results from a large dust cloud in a spiral arm
or a molecular ring. In addition, they also found a broad 2.2$\mum$
valley between about $l=22^{\circ}$ and $26^{\circ}$, which
corresponds to the expected position of the 3-kpc ring. Again, the
dip of $A_\lambda/A_\Ks$ around $l=25^{\circ}$ may be
consistent with the location of the 3$\kpc$ ring.

The dip of the longitudinal distribution of $A_\lambda/A_\Ks$ at
$l\simeq-30^{\circ}$ appears shifted from the position of the Norma
arm at $l=-33^{\circ}$. But we note that the longitudinal position
of the tangent to the Norma arm is not precisely known and the
reported positions range from $l=-37^{\circ}$ to $-28^{\circ}$
[\citealt{Vallee08}; e.g. $l=-31^{\circ}$ \citep{Bloemen90},
$l=-29^{\circ}$ \citep{Hammersley94}, $l=-32^{\circ}$
\citep{Bronfman08}]. This probably also explains why the dip of the
longitudinal distribution of $A_\lambda/A_\Ks$ at $l=48^{\circ}$
does not precisely match the tangent direction to the
Sagittarius-Carina arm at ``$l=50^{\circ}$'' (for which the reported
positions range from $l=46^{\circ}$ to $56^{\circ}$,
\citealt{Vallee08}).

Why does $A_\lambda/A_\Ks$ reach its minimum in the
Galactic spiral arm regions? A simple explanation would be grain
growth: in spiral galaxies, interstellar gas and dust are
concentrated in the inner edges of spiral arms (probably caused by
the spiral density wave; \citealt{Greenberg70}). \citet{Greenberg95}
argued that as the interarm matter encounters the density wave
potential minimum, the gas is compressed by the density wave shock
(with a speed of a few tens km\,s$^{-1}$;
\citealt{Roberts69, Roberts75, Roberts79}) at the inner edge of the
spiral arm. This would lead not only to a larger number density of
gas and dust, but also an increase of grain size by
accretion and coagulation. As shown in Figure\,7, if the
dust in the spiral arm regions (where giant molecular clouds are
strongly concentrated) grows to $a \simali$0.2--0.3$\mum$, the
wavelength dependence of $A_\lambda/A_\Ks$ in the wavelength range
of 2--8$\mum$ becomes steeper (i.e. smaller) than that of dust of
radii $a=0.1\mum$ (mean size for the dust in the diffuse ISM) for
amorphous silicate, amorphous carbon, and graphite. This explains
why the $A_\lambda/A_\Ks$ values are small in the Galactic spiral
arms for all four IRAC bands.

\paragraph{Relationship with the dust IR emission:}
%
In Figure\,\ref{fig8} we plot the longitudinal distribution of the
Galactic 240$\mum$ emission
from the DIRBE Galactic Plane Maps\footnote{%
   The data are available on {\sf
   http://lambda.gsfc.nasa.gov/product/cobe/dirbe\_gpm\_data\_get.cfm}.
   }
with $A_{[3.6]}/A_\Ks$, as well as the locations of the spiral arms
\citep{Vallee08}. We see that most of the dips of the
$A_{[3.6]}/A_\Ks$ profile coincide with the peaks of the 240$\mum$
emission (e.g. the features around $l=-55^{\circ},-49^{\circ},
-29^{\circ}, -23^{\circ}, -8^{\circ}, 25^{\circ}, 31^{\circ},
38^{\circ}, 49^{\circ}$).
\citet{Drimmel00} found that the features
in the 240$\mum$ emission profile from dust
can be identified with spiral arm tangents.
Similarly, most of the dips of the extinction profile
can be understood in terms of the locations of
the spiral arms (e.g. the 240$\mum$ emission peak
and the $A_{[3.6]}/A_\Ks$ dip at $l=-49^{\circ}$ vs.
the Crux-Scutum arm,
the $l=-29^{\circ}$ feature vs. the Norma arm,
the $l=31^{\circ}$ feature vs. the Scutum-Crux arm, the
$l=49^{\circ}$ feature vs. the Sagittarius-Carina arm;
the features around $l=-23^{\circ}$ and $l=25^{\circ}$
vs. the 3$\kpc$ ring or arm):
in the spiral arm regions which are rich in star-forming giant
molecular clouds and HII regions, not only is the dust size
relatively larger because of accretion and coagulational growth
(which explains the coincidence of the dips of $A_\lambda/A_\Ks$
with the locations of the spiral arms),
but the dust number density is also higher
because of the gas and dust concentration in the inner
edges of spiral arms caused by the spiral density wave,
and the starlight intensity which illuminates
the dust causing it to emit at 240$\mum$ is also
higher because of the star formation activity occurring
in the arm. Both the increased dust concentration
and starlight intensity would result in
a higher 240$\mum$ emission.

\paragraph{The Galactic bulge region:}
%
In all four IRAC bands, the extinction ratios $A_\lambda/A_\Ks$
derived from red giants decrease for the sightlines toward the bulge
region and reaches their minimum values near $l=0^{\circ}$, while it
is the opposite for that derived from the RCG samples. This
difference is probably due to the limited distance which the RCG
samples can probe, as discussed earlier in \S4.2.2. In addition, the
extinction is not simply a linear function of distance because of
the nonuniform distribution (i.e. grain density, size, and
composition)
of interstellar dust along these directions.\footnote{%
   For example, it has been observationally
           demonstrated that both the ratio of
           the visual extinction ($A_V$) to the 9.7$\mum$
           Si--O optical depth ($\tausil$)
           and the ratio of $A_V$ to the 3.4$\mum$
           C--H optical depth ($\tauahc$) show
           considerable variations from the local diffuse ISM
           to the GC (see \citealt{Gao09}).
           \citet{Sandford95} explained this
           by assuming that the abundance of
           the C--H carrier (relative to other dust components)
           gradually increases from the local ISM toward the GC.
           Using the RCG stars in the GC within $|l|<2^{\circ}$
           as a probe, \citet{Nishiyama06a} found that
           the interstellar extinction of these areas is indeed
           highly nonuniform.
           Also the conventional linear
           relation between extinction and distance of
           $A_V/d\approx 1.8\magni\,{\rm kpc}^{-1}$
           was obtained for stars within $\simali$1000\,pc
           of the Sun and within $\simali$100\,pc of
           the galactic plane \citep{Spitzer78}.
           Toward the galactic bulge, in view of the large scale
           structure of the ISM, the small-scale patchy distribution
           of dust seems well established
           (see \citealt{Spitzer78}).
   }
Thus the extinction probed by red giants and RCG stars
in this work may be very different for the sightlines
toward the bulge region.

The decreasing trend of $A_\lambda/A_\Ks$ toward the GC suggests
that the extinction law toward the GC may be different from that of
the local diffuse ISM or star-forming regions. Recently, using the
bulge RCG stars, \citet{Nishiyama09} obtained the
$\simali$1.2--8$\mum$ interstellar extinction law toward the GC.
They found an appreciably steeper extinction law compared with that
of the off-cloud regions in the Galactic plane of
\citet{Indebetouw05} and the star-forming regions of
\citet{Flaherty07}.

\citet{Jiang06} studied the ISOGAL fields using red giants and found
no clear structural distribution of the 7$\mum$ extinction along
Galactic longitude. We re-analyzed the extinction around 7$\mum$ by
smoothing the distribution through a 3-degree bin. The results
(based on the ISO/SWS 6.7$\mum$ LW2, 6.8$\mum$ LW5, and 7.7 $\mum$
LW6 data) are shown in Figure\,\ref{fig9}. The smoothed, re-analyzed
data appear to display a longitude variation of $A_{[7\mum]}/A_\Ks$,
with the local minimum at the GC as the most notable feature. In
particular, the extinction ratio $A_{7.7\mum}/A_\Ks$ shows a profile
similar to that derived from the RG1 samples.

\section{Conclusions}\label{conclusions}
Using red giants and red clump giants as tracers, we have derived
$A_\lambda/A_\Ks$, the extinction (relative to the 2MASS $\Ks$ band)
in the four IRAC bands, [3.6], [4.5], [5.8] and [8.0]$\mum$ for 131
GLIPMSE fields along the Galactic plane within $|l|\leq65^{\rm o}$,
based on the data obtained from the \textit{Spitzer}/GLIPMSE Legacy
Program and the 2MASS Survey project.
The principal results of this paper are the following:
\begin{enumerate}
\item The mean extinction in the IRAC bands
      (normalized to the 2MASS $\Ks$ band),
      $A_{[3.6]}/A_\Ks\approx0.63\pm0.01$,
      $A_{[4.5]}/A_\Ks\approx0.57\pm0.03$,
      $A_{[5.8]}/A_\Ks\approx0.49\pm0.01$, and
      $A_{[8.0]}/A_\Ks\approx0.55\pm0.03$,
      exhibits little variation with wavelength
      and lacks the minimum at $\simali$7$\mum$
      predicted from the standard interstellar grain model for $R_V=3.1$.
      This is consistent within errors
      with previous observational determinations
      based on {\it ISO} and {\it Spitzer} data
      and with that predicted from the grain model
      for $R_V=5.5$ of \citet{Weingartner01}.
\item The wavelength dependence of interstellar extinction in the mid-IR
      varies from one sightline to another, suggesting that there may not
      exist a ``universal'' IR extinction law.

\item There exist systematic variations of extinction with Galactic
      longitude which appears to correlate with the locations of spiral
      arms and with the variation of the 240$\mum$ dust emission.
      This can be understood in terms of larger grain sizes
      (arising from coagulational growth),
      enhanced dust concentration, and higher starlight intensities
      in the spiral arm regions.
\end{enumerate}

\acknowledgments{We thank the anonymous referee for his/her very
helpful comments. We thank the referee and Dr. Robert Benjamin for
providing us the extinction vs. CO emission data.
We thank Drs. Y.-Q. Lou and J.P. Vall\'ee for fruitful discussions.
This work is supported by the NSFC grants No.\,10603001 and 10973004,
the grant 2007CB815406. AL is supported in part by
the Spitzer Cycle 3 GO program P30403,
the NASA/Herschel Cycle 0 Theory grant,
and NSF grant AST 07-07866.}

{\it Facilities:} \facility{\emph{Spitzer}(IRAC)}, \facility{2MASS}.


\clearpage

\begin{figure}
\hspace{-0.5in} \centering
\includegraphics[angle=0,width=3.5in]{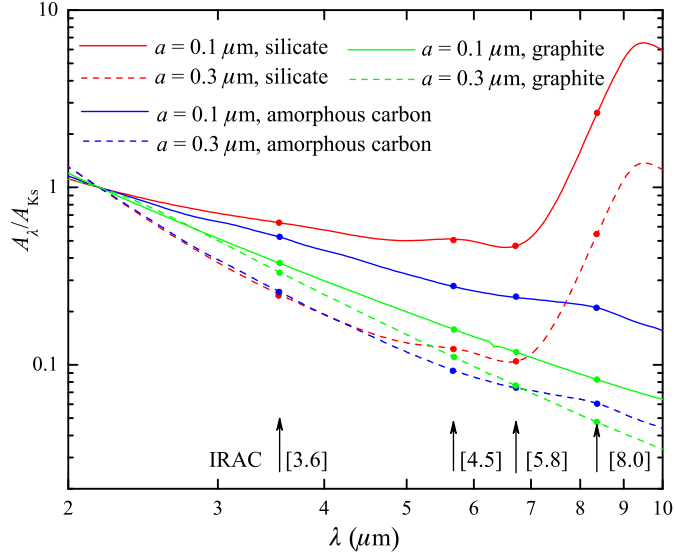}
\caption{\footnotesize
         The wavelength dependence of extinction (normalized
         in the $\Ks$ band) $A_{\lambda}/A_\Ks$ of spherical
         grains of radii $a=0.1\mum$ and 0.3$\mum$ of
         amorphous silicate, graphite, or amorphous carbon
         composition. The $A_{\lambda}/A_\Ks$ values
         in four IRAC bands are marked with filled circles.
         We see that if the dust in the Galactic spiral arm
         regions grows to $a\sim 0.3\mum$,
         one would expect a steeper extinction law
         (i.e. smaller $A_{\lambda}/A_\Ks$ ratios)
         than that of the diffuse ISM for which the mean dust
         size is $\simali$0.1$\mum$.
         \label{fig7}}
\end{figure}

\begin{figure}
\centering
\includegraphics[angle=0,width=3.5in]{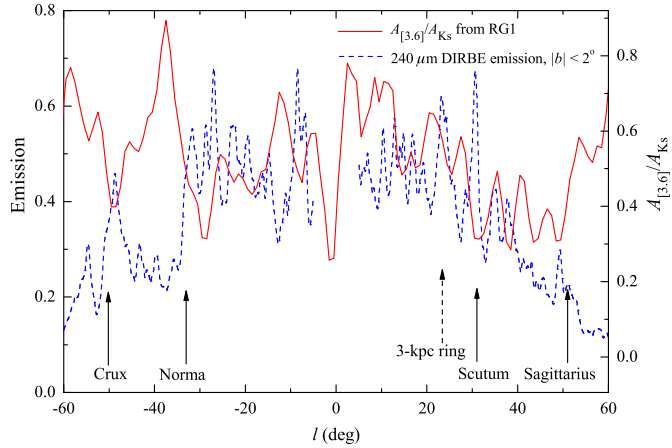}
\caption{\footnotesize
         Longitudinal profiles of $A_{[3.6]}/A_\Ks$ and the
         240$\mum$ dust emission from
         the DIRBE Galactic Plane Maps
         (averaged over the latitude interval $|b|<2^{\circ}$).
         Most of the dips on the $A_{[3.6]}/A_\Ks$ profile
         coincide with the 240$\mum$ emission peak.
         \label{fig8}}
\end{figure}

\begin{figure}
\centering
\includegraphics[angle=0,width=3.5in]{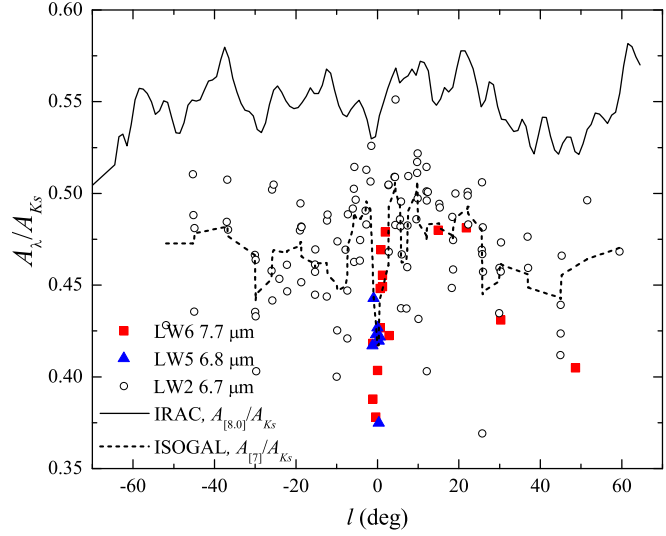}
\caption{\footnotesize
        Longitudinal profiles of $A_{[8.0]}/A_\Ks$
        (solid line) and $A_{[7]}/A_\Ks$ (dashed line).
        The latter was obtained from smoothing, re-analyzing
        the ISO LW2 (6.7$\mum$), LW5 (6.8$\mum$) and LW6 (7.7$\mum$) data of
        the ISOGAL fields \citep{Jiang06}.
        \label{fig9}}
\end{figure}

\begin{table}
\begin{center}
\caption{Average extinction (relative to $A_\Ks$) over all 131
GLIMPSE fields by different tracers
         \label{tbl-relext}} \vspace{0.2in}
\begin{tabular}{l|cccc}
\tableline \tableline Tracers & $A_{[3.6]}/A_\Ks$ &
$A_{[4.5]}/A_\Ks$ & $A_{[5.8]}/A_ \Ks$ & $A_{[8.0]}/A_\Ks$\\
\tableline
Red Giants (RG1, J) & 0.64$\pm$0.01 & 0.63$\pm$0.01 & 0.52$\pm$0.01 & 0.55$\pm$0.02 \\
\textit{Red Giants (RG2, \J)} & \textit{0.65$\pm$0.01} & \textit{0.63$\pm$0.02} & \textit{0.53$\pm$0.01} & \textit{0.55$\pm$0.02}\\
\textit{Red Giants (RG2, \HH)} & \textit{0.61$\pm$0.01} & \textit{0.59$\pm$0.02} & \textit{0.48$\pm$0.01} & \textit{0.50$\pm$0.02}\\
Red Clump Giants (RCG, J) & 0.61$\pm$0.01 & 0.51$\pm$0.04 & 0.45$\pm$0.04& 0.55$\pm$0.04\\

\tableline mean \tablenotemark{a} & 0.63$\pm$0.01 & 0.57$\pm$0.03 & 0.49$\pm$0.03& 0.55$\pm$0.03\\
\tableline Nishiyama09\tablenotemark{b} & 0.50$\pm$0.01 &
0.39$\pm$0.01 & 0.36$\pm$0.01 &
0.43$\pm$0.01\\Flaherty07\tablenotemark{c} & 0.631$\pm$0.005 &
0.53$\pm$0.01 & 0.48$\pm$0.01 & 0.49$\pm$0.01 \\
Indebetouw05\tablenotemark{d} & 0.56$\pm$0.06 & 0.43$\pm$0.08 & 0.43$\pm$0.10 & 0.43$\pm$0.10\\
\tableline
\end{tabular}
\tablenotetext{a}{The mean extinction is calculated from the RG1 and
RCG results.} \tablenotetext{b}{Extinction toward the GC
\citep{Nishiyama09}} \tablenotetext{c}{These are averaged results
over five star-forming regions \citep{Flaherty07}.}
\tablenotetext{d}{The uncertainties on $A_{x}/A_{\Ks}$ of
\citet{Indebetouw05} include the uncertainty of $A_{\rm
H}/A_{\Ks}$.}
\end{center}
\end{table}

\begin{table}
\begin{center}
\caption{Extinction (relative to $A_\Ks$) for the GLIMPSE
Observation Strategy Validation (OSV) field, with $283.8^{\circ}
\leq l \leq 284.6^{\circ}$ and $|b|<1^{\circ}$ \label{ext_OSV}}
\vspace{0.2in}
\begin{tabular}{l|cccc}
\tableline \tableline Tracers & $A_{[3.6]}/A_\Ks$ &
$A_{[4.5]}/A_\Ks$ & $A_{[5.8]}/A_ \Ks$ & $A_{[8.0]}/A_\Ks$\\
\tableline
Red Giants (RG1, J) & 0.609$\pm$0.002 & 0.574$\pm$0.004 & 0.478$\pm$0.004 & 0.474$\pm$0.005 \\
Red Giants (RG2, \J) & 0.612$\pm$0.002 & 0.585$\pm$0.004 & 0.487$\pm$0.004 & 0.483$\pm$0.005 \\
Red Giants (RG2, \HH) & 0.578$\pm$0.002 & 0.534$\pm$0.004 & 0.429$\pm$0.005 & 0.425$\pm$0.005 \\
Red Clump Giants (RCG, J) & 0.585$\pm$0.005 & 0.436$\pm$0.006 & 0.335$\pm$0.007& 0.40$\pm$0.01\\
\tableline
Indebetouw05\tablenotemark{a} & 0.57$\pm$0.05 & 0.43$\pm$0.07 & 0.41$\pm$0.07 & 0.37$\pm$0.07\\
Flaherty07 & 0.57$\pm$0.01 & 0.50$\pm$0.01 & 0.40$\pm$0.01 & 0.39$\pm$0.01 \\
\tableline
\end{tabular}
\tablenotetext{a}{Quoted values taken from \citet{Indebetouw05} are
for the off-cloud line of sight. They used red clump giants as
tracers and the $\J$ band as the comparison band, with
$A_\J/A_\Ks=2.5\pm0.2$.}
\end{center}
\end{table}

%
\begin{deluxetable}{p{1.7cm}|p{1.3cm}p{1.4cm}|p{1.5cm}p{1.5cm}p{1.5cm}p{1.6cm}|p{1.3cm}p{1.3cm}p{1.3cm}p{1.3cm}p{0.8cm}}
\rotate \tabletypesize{\scriptsize} \tablewidth{21.5cm}
\tablecaption{\footnotesize
        Extinction ratios $A_{\lambda}/A_\Ks$ of the 131 GLIMPSE
        fields\tablenotemark{a}\label{131tab}} \vspace{0.01in}
\startdata \hline \hline
               & \multicolumn{2}{|c|}{}      & \multicolumn{4}{c|}{Red Giants (RG1)} & \multicolumn{5}{c}{Red Clump Giants} \\
Field Name     & $l$\,(deg)    & $b$\,(deg)  & $A_{[3.6]}/A_\Ks$ & $A_{[4.5]}/A_\Ks$ & $A_{[5.8]}/A_\Ks$ & $A_{[8.0]}/A_\Ks$ & $A_{[3.6]}/A_\Ks$ & $A_{[4.5]}/A_\Ks$ & $A_{[5.8]}/A_\Ks$ & $A_{[8.0]}/A_\Ks$ & Max($\Ks$) \\
\hline
GLMIIC{\_}l000 & [0,1]         & [-2,+2]     & 0.651$\pm$0.001 & 0.630$\pm$0.001 & 0.541$\pm$0.001 & 0.550$\pm$0.001 & 0.70$\pm$0.01 & 0.83$\pm$0.02 & 0.56$\pm$0.02 & 0.78$\pm$0.02  &   11.5   \\
GLMIIC{\_}l001 & [1,2]         & [-2,+2]     & 0.667$\pm$0.001 & 0.605$\pm$0.001 & 0.526$\pm$0.001 & 0.555$\pm$0.001 & 0.68$\pm$0.01 & 0.78$\pm$0.01 & 0.57$\pm$0.01 & 0.79$\pm$0.01  &   11.5   \\
GLMIIC{\_}l002 & [2,3]         & [-1.5,+1.5] & 0.651$\pm$0.001 & 0.591$\pm$0.001 & 0.512$\pm$0.001 & 0.541$\pm$0.001 & 0.62$\pm$0.01 & 0.67$\pm$0.01 & 0.48$\pm$0.01 & 0.71$\pm$0.01  &   11.5   \\
GLMIIC{\_}l003 & [3,4]         & [-1.5,+1.5] & 0.681$\pm$0.001 & 0.642$\pm$0.001 & 0.547$\pm$0.001 & 0.578$\pm$0.001 & 0.59$\pm$0.01 & 0.57$\pm$0.01 & 0.46$\pm$0.01 & 0.66$\pm$0.01  &   11.5   \\
GLMIIC{\_}l004 & [4,5]         & [-1.5,+1.5] & 0.656$\pm$0.001 & 0.628$\pm$0.001 & 0.533$\pm$0.001 & 0.569$\pm$0.001 & 0.62$\pm$0.01 & 0.56$\pm$0.01 & 0.46$\pm$0.01 & 0.64$\pm$0.01  &   11.5   \\
GLMIIC{\_}l005 & [5,6]         & [-1,+1]     & 0.658$\pm$0.001 & 0.634$\pm$0.001 & 0.531$\pm$0.001 & 0.562$\pm$0.001 & 0.60$\pm$0.01 & 0.57$\pm$0.01 & 0.49$\pm$0.01 & 0.66$\pm$0.01  &   12.0   \\
GLMIIC{\_}l006 & [6,7]         & [-1,+1]     & 0.651$\pm$0.001 & 0.628$\pm$0.001 & 0.523$\pm$0.001 & 0.551$\pm$0.001 & 0.59$\pm$0.01 & 0.54$\pm$0.01 & 0.47$\pm$0.01 & 0.63$\pm$0.02  &   12.0   \\
GLMIIC{\_}l007 & [7,8]         & [-1,+1]     & 0.665$\pm$0.001 & 0.644$\pm$0.001 & 0.537$\pm$0.001 & 0.576$\pm$0.001 & 0.57$\pm$0.01 & 0.54$\pm$0.01 & 0.44$\pm$0.01 & 0.59$\pm$0.01  &   12.0   \\
GLMIIC{\_}l008 & [8,9]         & [-1,+1]     & 0.658$\pm$0.001 & 0.632$\pm$0.001 & 0.528$\pm$0.001 & 0.557$\pm$0.001 & 0.61$\pm$0.01 & 0.57$\pm$0.01 & 0.50$\pm$0.01 & 0.62$\pm$0.01  &   12.0   \\
GLMXC{\_}l009  & [9,10]        & [-1,+1]     & 0.663$\pm$0.001 & 0.640$\pm$0.001 & 0.532$\pm$0.001 & 0.568$\pm$0.001 & 0.59$\pm$0.01 & 0.54$\pm$0.01 & 0.46$\pm$0.01 & 0.60$\pm$0.01  &   11.5   \\
GLMIC{\_}l010  & [10,11]       & [-1,+1]     & 0.645$\pm$0.001 & 0.620$\pm$0.001 & 0.517$\pm$0.001 & 0.560$\pm$0.001 & 0.61$\pm$0.01 & 0.51$\pm$0.01 & 0.46$\pm$0.01 & 0.56$\pm$0.01  &   11.5   \\
GLMIC{\_}l011  & [11,12]       & [-1,+1]     & 0.671$\pm$0.001 & 0.669$\pm$0.001 & 0.552$\pm$0.001 & 0.587$\pm$0.001 & 0.58$\pm$0.01 & 0.46$\pm$0.01 & 0.43$\pm$0.01 & 0.56$\pm$0.01  &   11.5   \\
GLMIC{\_}l012  & [12,13]       & [-1,+1]     & 0.650$\pm$0.001 & 0.633$\pm$0.001 & 0.525$\pm$0.001 & 0.553$\pm$0.001 & 0.59$\pm$0.01 & 0.47$\pm$0.01 & 0.40$\pm$0.01 & 0.57$\pm$0.01  &   12.0   \\
GLMIC{\_}l013  & [13,14]       & [-1,+1]     & 0.648$\pm$0.001 & 0.618$\pm$0.001 & 0.518$\pm$0.001 & 0.554$\pm$0.001 & 0.60$\pm$0.01 & 0.51$\pm$0.01 & 0.46$\pm$0.01 & 0.58$\pm$0.01  &   12.0   \\
GLMIC{\_}l014  & [14,15]       & [-1,+1]     & 0.636$\pm$0.001 & 0.602$\pm$0.001 & 0.495$\pm$0.001 & 0.529$\pm$0.001 & 0.62$\pm$0.01 & 0.58$\pm$0.01 & 0.50$\pm$0.01 & 0.61$\pm$0.01  &   12.0   \\
GLMIC{\_}l015  & [15,16]       & [-1,+1]     & 0.646$\pm$0.001 & 0.630$\pm$0.001 & 0.521$\pm$0.001 & 0.559$\pm$0.001 & 0.62$\pm$0.01 & 0.56$\pm$0.01 & 0.49$\pm$0.01 & 0.61$\pm$0.01  &   12.0   \\
GLMIC{\_}l016  & [16,17]       & [-1,+1]     & 0.648$\pm$0.001 & 0.634$\pm$0.001 & 0.523$\pm$0.001 & 0.555$\pm$0.001 & 0.60$\pm$0.01 & 0.49$\pm$0.01 & 0.44$\pm$0.01 & 0.56$\pm$0.01  &   12.0   \\
GLMIC{\_}l017  & [17,18]       & [-1,+1]     & 0.643$\pm$0.001 & 0.627$\pm$0.001 & 0.517$\pm$0.001 & 0.547$\pm$0.001 & 0.60$\pm$0.01 & 0.51$\pm$0.01 & 0.44$\pm$0.01 & 0.57$\pm$0.01  &   12.0   \\
GLMIC{\_}l018  & [18,19]       & [-1,+1]     & 0.640$\pm$0.001 & 0.624$\pm$0.001 & 0.519$\pm$0.001 & 0.555$\pm$0.001 & 0.62$\pm$0.01 & 0.56$\pm$0.01 & 0.50$\pm$0.01 & 0.60$\pm$0.01  &   12.0   \\
GLMIC{\_}l019  & [19,20]       & [-1,+1]     & 0.653$\pm$0.001 & 0.634$\pm$0.001 & 0.532$\pm$0.001 & 0.568$\pm$0.001 & 0.62$\pm$0.01 & 0.56$\pm$0.01 & 0.50$\pm$0.01 & 0.60$\pm$0.01  &   12.0   \\
GLMIC{\_}l020  & [20,21]       & [-1,+1]     & 0.655$\pm$0.001 & 0.660$\pm$0.001 & 0.546$\pm$0.001 & 0.585$\pm$0.001 & 0.61$\pm$0.01 & 0.56$\pm$0.01 & 0.47$\pm$0.01 & 0.59$\pm$0.01  &   12.0   \\
GLMIC{\_}l021  & [21,22]       & [-1,+1]     & 0.652$\pm$0.001 & 0.650$\pm$0.001 & 0.539$\pm$0.001 & 0.580$\pm$0.001 & 0.61$\pm$0.01 & 0.53$\pm$0.01 & 0.46$\pm$0.01 & 0.57$\pm$0.01  &   12.0   \\
GLMIC{\_}l022  & [22,23]       & [-1,+1]     & 0.650$\pm$0.001 & 0.628$\pm$0.001 & 0.528$\pm$0.001 & 0.565$\pm$0.001 & 0.61$\pm$0.01 & 0.53$\pm$0.01 & 0.47$\pm$0.01 & 0.59$\pm$0.01  &   12.0   \\
GLMIC{\_}l023  & [23,24]       & [-1,+1]     & 0.647$\pm$0.001 & 0.627$\pm$0.001 & 0.528$\pm$0.001 & 0.569$\pm$0.001 & 0.61$\pm$0.01 & 0.53$\pm$0.01 & 0.47$\pm$0.01 & 0.58$\pm$0.01  &   12.0   \\
GLMIC{\_}l024  & [24,25]       & [-1,+1]     & 0.644$\pm$0.001 & 0.620$\pm$0.001 & 0.522$\pm$0.001 & 0.560$\pm$0.001 & 0.65$\pm$0.01 & 0.56$\pm$0.01 & 0.49$\pm$0.01 & 0.59$\pm$0.02  &   12.0   \\
GLMIC{\_}l025  & [25,26]       & [-1,+1]     & 0.642$\pm$0.001 & 0.617$\pm$0.001 & 0.513$\pm$0.001 & 0.541$\pm$0.001 & 0.59$\pm$0.01 & 0.51$\pm$0.01 & 0.45$\pm$0.01 & 0.57$\pm$0.01  &   12.0   \\
GLMIC{\_}l026  & [26,27]       & [-1,+1]     & 0.647$\pm$0.001 & 0.634$\pm$0.001 & 0.524$\pm$0.001 & 0.544$\pm$0.001 & 0.60$\pm$0.01 & 0.55$\pm$0.01 & 0.47$\pm$0.01 & 0.61$\pm$0.01  &   12.0   \\
GLMIC{\_}l027  & [27,28]       & [-1,+1]     & 0.656$\pm$0.001 & 0.650$\pm$0.001 & 0.544$\pm$0.001 & 0.581$\pm$0.001 & 0.64$\pm$0.01 & 0.58$\pm$0.01 & 0.51$\pm$0.01 & 0.64$\pm$0.01  &   12.0   \\
GLMIC{\_}l028  & [28,29]       & [-1,+1]     & 0.650$\pm$0.001 & 0.633$\pm$0.001 & 0.533$\pm$0.001 & 0.565$\pm$0.001 & 0.62$\pm$0.01 & 0.54$\pm$0.01 & 0.48$\pm$0.01 & 0.60$\pm$0.01  &   12.0   \\
GLMIC{\_}l029  & [29,30]       & [-1,+1]     & 0.634$\pm$0.001 & 0.609$\pm$0.001 & 0.511$\pm$0.001 & 0.536$\pm$0.001 & 0.60$\pm$0.01 & 0.53$\pm$0.01 & 0.47$\pm$0.01 & 0.62$\pm$0.01  &   12.0   \\
GLMIC{\_}l030  & [30,31]       & [-1,+1]     & 0.627$\pm$0.001 & 0.599$\pm$0.001 & 0.503$\pm$0.001 & 0.534$\pm$0.001 & 0.65$\pm$0.01 & 0.54$\pm$0.01 & 0.45$\pm$0.01 & 0.57$\pm$0.01  &   12.0   \\
GLMIC{\_}l031  & [31,32]       & [-1,+1]     & 0.630$\pm$0.001 & 0.607$\pm$0.001 & 0.506$\pm$0.001 & 0.533$\pm$0.001 & 0.61$\pm$0.01 & 0.51$\pm$0.01 & 0.42$\pm$0.01 & 0.55$\pm$0.01  &   12.0   \\
GLMIC{\_}l032  & [32,33]       & [-1,+1]     & 0.634$\pm$0.001 & 0.615$\pm$0.001 & 0.512$\pm$0.001 & 0.540$\pm$0.001 & 0.64$\pm$0.01 & 0.63$\pm$0.01 & 0.53$\pm$0.01 & 0.63$\pm$0.01  &   12.0   \\
GLMIC{\_}l033  & [33,34]       & [-1,+1]     & 0.632$\pm$0.001 & 0.611$\pm$0.001 & 0.510$\pm$0.001 & 0.530$\pm$0.001 & 0.62$\pm$0.01 & 0.58$\pm$0.01 & 0.43$\pm$0.01 & 0.50$\pm$0.01  &   13.0   \\
GLMIC{\_}l034  & [34,35]       & [-1,+1]     & 0.641$\pm$0.001 & 0.614$\pm$0.001 & 0.514$\pm$0.001 & 0.531$\pm$0.001 & 0.60$\pm$0.01 & 0.49$\pm$0.01 & 0.44$\pm$0.01 & 0.52$\pm$0.01  &   12.0   \\
\cline{1-12} \multicolumn{12}{c}{} \\ \hline \hline
               & \multicolumn{2}{|c|}{}      & \multicolumn{4}{c|}{Red Giants (RG1)} & \multicolumn{5}{c}{Red Clump Giants} \\
Field Name     & $l$\,(deg)    & $b$\,(deg)  & $A_{[3.6]}/A_\Ks$ & $A_{[4.5]}/A_\Ks$ & $A_{[5.8]}/A_\Ks$ & $A_{[8.0]}/A_\Ks$ & $A_{[3.6]}/A_\Ks$ & $A_{[4.5]}/A_\Ks$ & $A_{[5.8]}/A_\Ks$ & $A_{[8.0]}/A_\Ks$ & Max($\Ks$) \\
\hline
GLMIC{\_}l035  & [35,36]       & [-1,+1]     & 0.643$\pm$0.001 & 0.634$\pm$0.001 & 0.526$\pm$0.001 & 0.536$\pm$0.001 & 0.60$\pm$0.01 & 0.48$\pm$0.01 & 0.42$\pm$0.01 & 0.51$\pm$0.01  &   12.0   \\
GLMIC{\_}l036  & [36,37]       & [-1,+1]     & 0.641$\pm$0.001 & 0.630$\pm$0.001 & 0.527$\pm$0.001 & 0.546$\pm$0.001 & 0.60$\pm$0.01 & 0.51$\pm$0.01 & 0.44$\pm$0.01 & 0.54$\pm$0.01  &   12.0   \\
GLMIC{\_}l037  & [37,38]       & [-1,+1]     & 0.625$\pm$0.001 & 0.601$\pm$0.001 & 0.501$\pm$0.001 & 0.515$\pm$0.001 & 0.62$\pm$0.01 & 0.60$\pm$0.01 & 0.50$\pm$0.01 & 0.56$\pm$0.02  &   12.5   \\
GLMIC{\_}l038  & [38,39]       & [-1,+1]     & 0.628$\pm$0.001 & 0.606$\pm$0.001 & 0.502$\pm$0.001 & 0.511$\pm$0.001 & 0.62$\pm$0.01 & 0.55$\pm$0.01 & 0.46$\pm$0.01 & 0.53$\pm$0.01  &   12.0   \\
GLMIC{\_}l039  & [39,40]       & [-1,+1]     & 0.637$\pm$0.001 & 0.632$\pm$0.001 & 0.518$\pm$0.001 & 0.533$\pm$0.001 & 0.59$\pm$0.01 & 0.46$\pm$0.01 & 0.41$\pm$0.01 & 0.52$\pm$0.01  &   12.0   \\
GLMIC{\_}l040  & [40,41]       & [-1,+1]     & 0.646$\pm$0.001 & 0.650$\pm$0.001 & 0.534$\pm$0.001 & 0.546$\pm$0.001 & 0.62$\pm$0.01 & 0.55$\pm$0.01 & 0.47$\pm$0.01 & 0.54$\pm$0.01  &   12.0   \\
GLMIC{\_}l041  & [41,42]       & [-1,+1]     & 0.644$\pm$0.001 & 0.641$\pm$0.001 & 0.530$\pm$0.001 & 0.548$\pm$0.001 & 0.61$\pm$0.01 & 0.53$\pm$0.01 & 0.46$\pm$0.01 & 0.53$\pm$0.01  &   12.0   \\
GLMIC{\_}l042  & [42,43]       & [-1,+1]     & 0.633$\pm$0.001 & 0.616$\pm$0.001 & 0.515$\pm$0.001 & 0.535$\pm$0.001 & 0.60$\pm$0.01 & 0.48$\pm$0.01 & 0.42$\pm$0.01 & 0.51$\pm$0.01  &   12.0   \\
GLMIC{\_}l043  & [43,44]       & [-1,+1]     & 0.628$\pm$0.001 & 0.598$\pm$0.001 & 0.499$\pm$0.001 & 0.514$\pm$0.001 & 0.59$\pm$0.01 & 0.50$\pm$0.01 & 0.44$\pm$0.01 & 0.53$\pm$0.01  &   12.0   \\
GLMIC{\_}l044  & [44,45]       & [-1,+1]     & 0.634$\pm$0.001 & 0.611$\pm$0.001 & 0.513$\pm$0.001 & 0.531$\pm$0.001 & 0.62$\pm$0.01 & 0.54$\pm$0.01 & 0.44$\pm$0.01 & 0.54$\pm$0.01  &   12.0   \\
GLMIC{\_}l045  & [45,46]       & [-1,+1]     & 0.635$\pm$0.001 & 0.607$\pm$0.001 & 0.511$\pm$0.001 & 0.525$\pm$0.001 & 0.62$\pm$0.01 & 0.58$\pm$0.01 & 0.48$\pm$0.01 & 0.57$\pm$0.01  &   12.0   \\
GLMIC{\_}l046  & [46,47]       & [-1,+1]     & 0.638$\pm$0.001 & 0.619$\pm$0.001 & 0.519$\pm$0.001 & 0.534$\pm$0.001 & 0.61$\pm$0.01 & 0.52$\pm$0.01 & 0.44$\pm$0.01 & 0.51$\pm$0.01  &   12.0   \\
GLMIC{\_}l047  & [47,48]       & [-1,+1]     & 0.636$\pm$0.001 & 0.625$\pm$0.002 & 0.523$\pm$0.001 & 0.534$\pm$0.001 & 0.60$\pm$0.01 & 0.53$\pm$0.01 & 0.43$\pm$0.01 & 0.51$\pm$0.01  &   12.0   \\
GLMIC{\_}l048  & [48,49]       & [-1,+1]     & 0.632$\pm$0.001 & 0.615$\pm$0.001 & 0.515$\pm$0.001 & 0.527$\pm$0.002 & 0.62$\pm$0.01 & 0.52$\pm$0.01 & 0.46$\pm$0.01 & 0.51$\pm$0.01  &   12.0   \\
GLMIC{\_}l049  & [49,50]       & [-1,+1]     & 0.627$\pm$0.001 & 0.605$\pm$0.002 & 0.507$\pm$0.001 & 0.518$\pm$0.002 & 0.60$\pm$0.01 & 0.49$\pm$0.01 & 0.43$\pm$0.01 & 0.51$\pm$0.01  &   12.0   \\
GLMIC{\_}l050  & [50,51]       & [-1,+1]     & 0.639$\pm$0.001 & 0.624$\pm$0.002 & 0.524$\pm$0.001 & 0.535$\pm$0.001 & 0.59$\pm$0.01 & 0.46$\pm$0.01 & 0.39$\pm$0.01 & 0.48$\pm$0.01  &   12.0   \\
GLMIC{\_}l051  & [51,52]       & [-1,+1]     & 0.640$\pm$0.001 & 0.621$\pm$0.001 & 0.523$\pm$0.001 & 0.533$\pm$0.001 & 0.60$\pm$0.01 & 0.47$\pm$0.01 & 0.40$\pm$0.01 & 0.48$\pm$0.01  &   12.0   \\
GLMIC{\_}l052  & [52,53]       & [-1,+1]     & 0.643$\pm$0.001 & 0.620$\pm$0.001 & 0.523$\pm$0.001 & 0.534$\pm$0.001 & 0.60$\pm$0.01 & 0.52$\pm$0.01 & 0.45$\pm$0.01 & 0.53$\pm$0.01  &   12.0   \\
GLMIC{\_}l053  & [53,54]       & [-1,+1]     & 0.646$\pm$0.001 & 0.632$\pm$0.002 & 0.533$\pm$0.002 & 0.539$\pm$0.002 & 0.61$\pm$0.01 & 0.54$\pm$0.01 & 0.46$\pm$0.01 & 0.53$\pm$0.01  &   12.0   \\
GLMIC{\_}l054  & [54,55]       & [-1,+1]     & 0.648$\pm$0.001 & 0.652$\pm$0.002 & 0.551$\pm$0.002 & 0.558$\pm$0.002 & 0.66$\pm$0.01 & 0.64$\pm$0.01 & 0.54$\pm$0.01 & 0.57$\pm$0.01  &   12.0   \\
GLMIC{\_}l055  & [55,56]       & [-1,+1]     & 0.639$\pm$0.001 & 0.617$\pm$0.002 & 0.528$\pm$0.002 & 0.535$\pm$0.002 & 0.62$\pm$0.01 & 0.53$\pm$0.01 & 0.40$\pm$0.01 & 0.50$\pm$0.01  &   12.5   \\
GLMIC{\_}l056  & [56,57]       & [-1,+1]     & 0.645$\pm$0.001 & 0.621$\pm$0.002 & 0.530$\pm$0.002 & 0.540$\pm$0.002 & 0.63$\pm$0.01 & 0.53$\pm$0.01 & 0.44$\pm$0.01 & 0.52$\pm$0.01  &   12.5   \\
GLMIC{\_}l057  & [57,58]       & [-1,+1]     & 0.642$\pm$0.001 & 0.619$\pm$0.002 & 0.530$\pm$0.002 & 0.538$\pm$0.002 & 0.62$\pm$0.01 & 0.55$\pm$0.01 & 0.48$\pm$0.01 & 0.53$\pm$0.01  &   12.0   \\
GLMIC{\_}l058  & [58,59]       & [-1,+1]     & 0.647$\pm$0.001 & 0.621$\pm$0.002 & 0.535$\pm$0.002 & 0.542$\pm$0.002 & 0.62$\pm$0.01 & 0.52$\pm$0.01 & 0.45$\pm$0.01 & 0.50$\pm$0.01  &   12.0   \\
GLMIC{\_}l059  & [59,60]       & [-1,+1]     & 0.644$\pm$0.001 & 0.633$\pm$0.002 & 0.543$\pm$0.002 & 0.548$\pm$0.002 & 0.60$\pm$0.01 & 0.51$\pm$0.01 & 0.45$\pm$0.01 & 0.51$\pm$0.01  &   12.0   \\
GLMIC{\_}l060  & [60,61]       & [-1,+1]     & 0.668$\pm$0.002 & 0.703$\pm$0.003 & 0.599$\pm$0.003 & 0.590$\pm$0.003 & 0.64$\pm$0.01 & 0.58$\pm$0.01 & 0.49$\pm$0.01 & 0.55$\pm$0.01  &   12.0   \\
GLMIC{\_}l061  & [61,62]       & [-1,+1]     & 0.665$\pm$0.002 & 0.691$\pm$0.003 & 0.585$\pm$0.003 & 0.589$\pm$0.003 & 0.63$\pm$0.01 & 0.56$\pm$0.01 & 0.48$\pm$0.01 & 0.52$\pm$0.01  &   12.0   \\
GLMIC{\_}l062  & [62,63]       & [-1,+1]     & 0.666$\pm$0.002 & 0.691$\pm$0.003 & 0.584$\pm$0.003 & 0.581$\pm$0.003 & 0.61$\pm$0.01 & 0.52$\pm$0.01 & 0.43$\pm$0.01 & 0.50$\pm$0.01  &   12.0   \\
GLMIC{\_}l063  & [63,64]       & [-1,+1]     & 0.666$\pm$0.002 & 0.682$\pm$0.003 & 0.580$\pm$0.003 & 0.577$\pm$0.003 & 0.62$\pm$0.01 & 0.53$\pm$0.01 & 0.48$\pm$0.01 & 0.51$\pm$0.01  &   12.0   \\
GLMIC{\_}l064  & [64,65]       & [-1,+1]     & 0.668$\pm$0.002 & 0.693$\pm$0.003 & 0.602$\pm$0.004 & 0.583$\pm$0.003 & 0.64$\pm$0.01 & 0.57$\pm$0.01 & 0.47$\pm$0.01 & 0.53$\pm$0.01  &   12.0   \\
GLMIC{\_}l283  & [283.8,284.6] & [-1,+1]     & 0.609$\pm$0.002 & 0.574$\pm$0.004 & 0.478$\pm$0.004 & 0.474$\pm$0.005 & 0.59$\pm$0.01 & 0.44$\pm$0.01 & 0.34$\pm$0.01 & 0.40$\pm$0.01  &   12.5   \\
GLMIC{\_}l295  & [295,296]     & [-1,+1]     & 0.646$\pm$0.002 & 0.662$\pm$0.004 & 0.568$\pm$0.004 & 0.548$\pm$0.004 & 0.62$\pm$0.01 & 0.53$\pm$0.01 & 0.37$\pm$0.01 & 0.49$\pm$0.01  &   12.5   \\
GLMIC{\_}l296  & [296,297]     & [-1,+1]     & 0.669$\pm$0.003 & 0.726$\pm$0.005 & 0.619$\pm$0.005 & 0.593$\pm$0.005 & 0.60$\pm$0.01 & 0.51$\pm$0.01 & 0.39$\pm$0.01 & 0.47$\pm$0.01  &   12.5   \\
GLMIC{\_}l297  & [297,298]     & [-1,+1]     & 0.644$\pm$0.002 & 0.648$\pm$0.003 & 0.544$\pm$0.003 & 0.531$\pm$0.003 & 0.59$\pm$0.01 & 0.48$\pm$0.01 & 0.39$\pm$0.01 & 0.47$\pm$0.01  &   12.0   \\
GLMIC{\_}l298  & [298,299]     & [-1,+1]     & 0.644$\pm$0.002 & 0.655$\pm$0.003 & 0.545$\pm$0.003 & 0.536$\pm$0.003 & 0.61$\pm$0.01 & 0.50$\pm$0.01 & 0.41$\pm$0.01 & 0.50$\pm$0.01  &   12.0   \\
\cline{1-12} \multicolumn{12}{c}{} \\ \hline \hline
               & \multicolumn{2}{|c|}{}      & \multicolumn{4}{c|}{Red Giants (RG1)} & \multicolumn{5}{c}{Red Clump Giants} \\
Field Name     & $l$\,(deg)    & $b$\,(deg)  & $A_{[3.6]}/A_\Ks$ & $A_{[4.5]}/A_\Ks$ & $A_{[5.8]}/A_\Ks$ & $A_{[8.0]}/A_\Ks$ & $A_{[3.6]}/A_\Ks$ & $A_{[4.5]}/A_\Ks$ & $A_{[5.8]}/A_\Ks$ & $A_{[8.0]}/A_\Ks$ & Max($\Ks$) \\
\hline
GLMIC{\_}l299  & [299,300]     & [-1,+1]     & 0.651$\pm$0.001 & 0.650$\pm$0.003 & 0.552$\pm$0.003 & 0.543$\pm$0.003 & 0.60$\pm$0.01 & 0.50$\pm$0.01 & 0.42$\pm$0.01 & 0.49$\pm$0.01  &   12.0   \\
GLMIC{\_}l300  & [300,301]     & [-1,+1]     & 0.660$\pm$0.001 & 0.662$\pm$0.002 & 0.561$\pm$0.002 & 0.558$\pm$0.002 & 0.58$\pm$0.01 & 0.48$\pm$0.01 & 0.40$\pm$0.01 & 0.47$\pm$0.01  &   12.0   \\
GLMIC{\_}l301  & [301,302]     & [-1,+1]     & 0.662$\pm$0.001 & 0.656$\pm$0.002 & 0.554$\pm$0.002 & 0.560$\pm$0.002 & 0.59$\pm$0.01 & 0.48$\pm$0.01 & 0.42$\pm$0.01 & 0.50$\pm$0.01  &   12.0   \\
GLMIC{\_}l302  & [302,303]     & [-1,+1]     & 0.651$\pm$0.001 & 0.630$\pm$0.002 & 0.535$\pm$0.002 & 0.549$\pm$0.002 & 0.61$\pm$0.01 & 0.55$\pm$0.01 & 0.45$\pm$0.01 & 0.53$\pm$0.01  &   12.0   \\
GLMIC{\_}l303  & [303,304]     & [-1,+1]     & 0.652$\pm$0.001 & 0.628$\pm$0.002 & 0.535$\pm$0.002 & 0.551$\pm$0.002 & 0.63$\pm$0.01 & 0.54$\pm$0.01 & 0.43$\pm$0.01 & 0.51$\pm$0.01  &   12.5   \\
GLMIC{\_}l304  & [304,305]     & [-1,+1]     & 0.648$\pm$0.001 & 0.628$\pm$0.001 & 0.532$\pm$0.001 & 0.552$\pm$0.002 & 0.62$\pm$0.01 & 0.53$\pm$0.01 & 0.46$\pm$0.01 & 0.54$\pm$0.01  &   12.0   \\
GLMIC{\_}l305  & [305,306]     & [-1,+1]     & 0.642$\pm$0.001 & 0.624$\pm$0.002 & 0.527$\pm$0.002 & 0.542$\pm$0.002 & 0.63$\pm$0.01 & 0.56$\pm$0.01 & 0.48$\pm$0.01 & 0.58$\pm$0.01  &   12.0   \\
GLMIC{\_}l306  & [306,307]     & [-1,+1]     & 0.647$\pm$0.001 & 0.607$\pm$0.002 & 0.521$\pm$0.001 & 0.542$\pm$0.002 & 0.63$\pm$0.01 & 0.58$\pm$0.01 & 0.47$\pm$0.01 & 0.54$\pm$0.01  &   12.5   \\
GLMIC{\_}l307  & [307,308]     & [-1,+1]     & 0.656$\pm$0.001 & 0.655$\pm$0.002 & 0.553$\pm$0.002 & 0.568$\pm$0.002 & 0.62$\pm$0.01 & 0.55$\pm$0.01 & 0.48$\pm$0.01 & 0.54$\pm$0.01  &   12.0   \\
GLMIC{\_}l308  & [308,309]     & [-1,+1]     & 0.648$\pm$0.001 & 0.634$\pm$0.001 & 0.533$\pm$0.001 & 0.553$\pm$0.002 & 0.60$\pm$0.01 & 0.50$\pm$0.01 & 0.43$\pm$0.01 & 0.51$\pm$0.01  &   12.0   \\
GLMIC{\_}l309  & [309,310]     & [-1,+1]     & 0.640$\pm$0.001 & 0.616$\pm$0.001 & 0.520$\pm$0.001 & 0.538$\pm$0.001 & 0.60$\pm$0.01 & 0.48$\pm$0.01 & 0.41$\pm$0.01 & 0.51$\pm$0.01  &   12.0   \\
GLMIC{\_}l310  & [310,311]     & [-1,+1]     & 0.637$\pm$0.001 & 0.616$\pm$0.001 & 0.517$\pm$0.001 & 0.531$\pm$0.001 & 0.61$\pm$0.01 & 0.47$\pm$0.01 & 0.41$\pm$0.01 & 0.53$\pm$0.01  &   12.0   \\
GLMIC{\_}l311  & [311,312]     & [-1,+1]     & 0.631$\pm$0.001 & 0.604$\pm$0.001 & 0.502$\pm$0.001 & 0.518$\pm$0.002 & 0.62$\pm$0.01 & 0.49$\pm$0.01 & 0.43$\pm$0.01 & 0.53$\pm$0.01  &   12.0   \\
GLMIC{\_}l312  & [312,313]     & [-1,+1]     & 0.640$\pm$0.001 & 0.616$\pm$0.001 & 0.517$\pm$0.001 & 0.536$\pm$0.001 & 0.61$\pm$0.01 & 0.47$\pm$0.01 & 0.41$\pm$0.01 & 0.52$\pm$0.01  &   12.0   \\
GLMIC{\_}l313  & [313,314]     & [-1,+1]     & 0.642$\pm$0.001 & 0.621$\pm$0.001 & 0.524$\pm$0.001 & 0.543$\pm$0.001 & 0.61$\pm$0.01 & 0.48$\pm$0.01 & 0.41$\pm$0.01 & 0.51$\pm$0.01  &   12.0   \\
GLMIC{\_}l314  & [314,315]     & [-1,+1]     & 0.649$\pm$0.001 & 0.621$\pm$0.001 & 0.525$\pm$0.001 & 0.552$\pm$0.001 & 0.61$\pm$0.01 & 0.49$\pm$0.01 & 0.43$\pm$0.01 & 0.53$\pm$0.01  &   12.0   \\
GLMIC{\_}l315  & [315,316]     & [-1,+1]     & 0.645$\pm$0.001 & 0.627$\pm$0.001 & 0.525$\pm$0.001 & 0.550$\pm$0.001 & 0.61$\pm$0.01 & 0.48$\pm$0.01 & 0.42$\pm$0.01 & 0.52$\pm$0.01  &   12.0   \\
GLMIC{\_}l316  & [316,317]     & [-1,+1]     & 0.641$\pm$0.001 & 0.616$\pm$0.001 & 0.518$\pm$0.001 & 0.548$\pm$0.001 & 0.60$\pm$0.01 & 0.48$\pm$0.01 & 0.41$\pm$0.01 & 0.51$\pm$0.01  &   12.0   \\
GLMIC{\_}l317  & [317,318]     & [-1,+1]     & 0.648$\pm$0.001 & 0.637$\pm$0.001 & 0.536$\pm$0.001 & 0.557$\pm$0.001 & 0.62$\pm$0.01 & 0.53$\pm$0.01 & 0.45$\pm$0.01 & 0.55$\pm$0.01  &   12.0   \\
GLMIC{\_}l318  & [318,319]     & [-1,+1]     & 0.648$\pm$0.001 & 0.625$\pm$0.001 & 0.532$\pm$0.001 & 0.562$\pm$0.001 & 0.63$\pm$0.01 & 0.54$\pm$0.01 & 0.46$\pm$0.01 & 0.55$\pm$0.01  &   12.0   \\
GLMIC{\_}l319  & [319,320]     & [-1,+1]     & 0.655$\pm$0.001 & 0.666$\pm$0.002 & 0.546$\pm$0.001 & 0.563$\pm$0.002 & 0.61$\pm$0.01 & 0.49$\pm$0.01 & 0.44$\pm$0.01 & 0.54$\pm$0.01  &   12.0   \\
GLMIC{\_}l320  & [320,321]     & [-1,+1]     & 0.650$\pm$0.001 & 0.651$\pm$0.001 & 0.540$\pm$0.001 & 0.556$\pm$0.001 & 0.60$\pm$0.01 & 0.48$\pm$0.01 & 0.43$\pm$0.01 & 0.55$\pm$0.01  &   12.0   \\
GLMIC{\_}l321  & [321,322]     & [-1,+1]     & 0.662$\pm$0.001 & 0.663$\pm$0.001 & 0.551$\pm$0.001 & 0.569$\pm$0.001 & 0.62$\pm$0.01 & 0.52$\pm$0.01 & 0.46$\pm$0.01 & 0.56$\pm$0.01  &   12.0   \\
GLMIC{\_}l322  & [322,323]     & [-1,+1]     & 0.672$\pm$0.001 & 0.691$\pm$0.001 & 0.571$\pm$0.001 & 0.592$\pm$0.001 & 0.62$\pm$0.01 & 0.52$\pm$0.01 & 0.45$\pm$0.01 & 0.55$\pm$0.01  &   12.0   \\
GLMIC{\_}l323  & [323,324]     & [-1,+1]     & 0.662$\pm$0.001 & 0.663$\pm$0.001 & 0.552$\pm$0.001 & 0.575$\pm$0.001 & 0.60$\pm$0.01 & 0.49$\pm$0.01 & 0.43$\pm$0.01 & 0.53$\pm$0.01  &   12.0   \\
GLMIC{\_}l324  & [324,325]     & [-1,+1]     & 0.650$\pm$0.001 & 0.646$\pm$0.001 & 0.535$\pm$0.001 & 0.555$\pm$0.001 & 0.61$\pm$0.01 & 0.50$\pm$0.01 & 0.43$\pm$0.01 & 0.54$\pm$0.01  &   12.0   \\
GLMIC{\_}l325  & [325,326]     & [-1,+1]     & 0.651$\pm$0.001 & 0.636$\pm$0.001 & 0.531$\pm$0.001 & 0.558$\pm$0.001 & 0.61$\pm$0.01 & 0.50$\pm$0.01 & 0.45$\pm$0.01 & 0.56$\pm$0.01  &   12.0   \\
GLMIC{\_}l326  & [326,327]     & [-1,+1]     & 0.645$\pm$0.001 & 0.624$\pm$0.001 & 0.523$\pm$0.001 & 0.544$\pm$0.001 & 0.61$\pm$0.01 & 0.48$\pm$0.01 & 0.43$\pm$0.01 & 0.55$\pm$0.01  &   12.0   \\
GLMIC{\_}l327  & [327,328]     & [-1,+1]     & 0.633$\pm$0.001 & 0.604$\pm$0.001 & 0.508$\pm$0.001 & 0.534$\pm$0.001 & 0.61$\pm$0.01 & 0.50$\pm$0.01 & 0.42$\pm$0.01 & 0.56$\pm$0.01  &   12.0   \\
GLMIC{\_}l328  & [328,329]     & [-1,+1]     & 0.641$\pm$0.001 & 0.615$\pm$0.001 & 0.518$\pm$0.001 & 0.546$\pm$0.001 & 0.60$\pm$0.01 & 0.48$\pm$0.01 & 0.42$\pm$0.01 & 0.54$\pm$0.01  &   12.0   \\
GLMIC{\_}l329  & [329,330]     & [-1,+1]     & 0.639$\pm$0.001 & 0.612$\pm$0.001 & 0.513$\pm$0.001 & 0.538$\pm$0.001 & 0.61$\pm$0.01 & 0.51$\pm$0.01 & 0.45$\pm$0.01 & 0.59$\pm$0.01  &   12.0   \\
GLMIC{\_}l330  & [330,331]     & [-1,+1]     & 0.631$\pm$0.001 & 0.611$\pm$0.001 & 0.505$\pm$0.001 & 0.532$\pm$0.001 & 0.59$\pm$0.01 & 0.47$\pm$0.01 & 0.40$\pm$0.01 & 0.53$\pm$0.01  &   12.0   \\
GLMIC{\_}l331  & [331,332]     & [-1,+1]     & 0.629$\pm$0.001 & 0.594$\pm$0.001 & 0.496$\pm$0.001 & 0.527$\pm$0.001 & 0.62$\pm$0.01 & 0.53$\pm$0.01 & 0.46$\pm$0.01 & 0.58$\pm$0.01  &   12.0   \\
GLMIC{\_}l332  & [332,333]     & [-1,+1]     & 0.639$\pm$0.001 & 0.605$\pm$0.001 & 0.506$\pm$0.001 & 0.533$\pm$0.001 & 0.61$\pm$0.01 & 0.51$\pm$0.01 & 0.47$\pm$0.01 & 0.59$\pm$0.01  &   12.0   \\
GLMIC{\_}l333  & [333,334]     & [-1,+1]     & 0.640$\pm$0.001 & 0.620$\pm$0.001 & 0.516$\pm$0.001 & 0.541$\pm$0.001 & 0.62$\pm$0.01 & 0.50$\pm$0.01 & 0.44$\pm$0.01 & 0.56$\pm$0.01  &   12.0   \\
\cline{1-12} \multicolumn{12}{c}{} \\ \hline \hline
               & \multicolumn{2}{|c|}{}      & \multicolumn{4}{c|}{Red Giants (RG1)} & \multicolumn{5}{c}{Red Clump Giants} \\
Field Name     & $l$\,(deg)    & $b$\,(deg)  & $A_{[3.6]}/A_\Ks$ & $A_{[4.5]}/A_\Ks$ & $A_{[5.8]}/A_\Ks$ & $A_{[8.0]}/A_\Ks$ & $A_{[3.6]}/A_\Ks$ & $A_{[4.5]}/A_\Ks$ & $A_{[5.8]}/A_\Ks$ & $A_{[8.0]}/A_\Ks$ & Max($\Ks$) \\
\hline
GLMIC{\_}l334  & [334,335]     & [-1,+1]     & 0.647$\pm$0.001 & 0.633$\pm$0.001 & 0.527$\pm$0.001 & 0.553$\pm$0.001 & 0.61$\pm$0.01 & 0.49$\pm$0.01 & 0.43$\pm$0.01 & 0.55$\pm$0.01  &   12.0   \\
GLMIC{\_}l335  & [335,336]     & [-1,+1]     & 0.648$\pm$0.001 & 0.639$\pm$0.001 & 0.531$\pm$0.001 & 0.558$\pm$0.001 & 0.61$\pm$0.01 & 0.49$\pm$0.01 & 0.43$\pm$0.01 & 0.56$\pm$0.01  &   12.0   \\
GLMIC{\_}l336  & [336,337]     & [-1,+1]     & 0.640$\pm$0.001 & 0.637$\pm$0.001 & 0.520$\pm$0.001 & 0.550$\pm$0.001 & 0.60$\pm$0.01 & 0.47$\pm$0.01 & 0.41$\pm$0.01 & 0.55$\pm$0.01  &   12.0   \\
GLMIC{\_}l337  & [337,338]     & [-1,+1]     & 0.641$\pm$0.001 & 0.628$\pm$0.001 & 0.517$\pm$0.001 & 0.545$\pm$0.001 & 0.61$\pm$0.01 & 0.50$\pm$0.01 & 0.44$\pm$0.01 & 0.57$\pm$0.01  &   12.0   \\
GLMIC{\_}l338  & [338,339]     & [-1,+1]     & 0.641$\pm$0.001 & 0.627$\pm$0.001 & 0.516$\pm$0.001 & 0.546$\pm$0.001 & 0.61$\pm$0.01 & 0.50$\pm$0.01 & 0.45$\pm$0.01 & 0.56$\pm$0.01  &   12.0   \\
GLMIC{\_}l339  & [339,340]     & [-1,+1]     & 0.644$\pm$0.001 & 0.629$\pm$0.001 & 0.519$\pm$0.001 & 0.539$\pm$0.001 & 0.57$\pm$0.01 & 0.44$\pm$0.01 & 0.39$\pm$0.01 & 0.56$\pm$0.01  &   12.0   \\
GLMIC{\_}l340  & [340,341]     & [-1,+1]     & 0.640$\pm$0.001 & 0.616$\pm$0.001 & 0.510$\pm$0.001 & 0.539$\pm$0.001 & 0.59$\pm$0.01 & 0.47$\pm$0.01 & 0.40$\pm$0.01 & 0.54$\pm$0.01  &   12.0   \\
GLMIC{\_}l341  & [341,342]     & [-1,+1]     & 0.636$\pm$0.001 & 0.619$\pm$0.001 & 0.513$\pm$0.001 & 0.545$\pm$0.001 & 0.61$\pm$0.01 & 0.47$\pm$0.01 & 0.42$\pm$0.01 & 0.56$\pm$0.01  &   12.0   \\
GLMIC{\_}l342  & [342,343]     & [-1,+1]     & 0.641$\pm$0.001 & 0.627$\pm$0.001 & 0.517$\pm$0.001 & 0.549$\pm$0.001 & 0.62$\pm$0.01 & 0.50$\pm$0.01 & 0.45$\pm$0.01 & 0.56$\pm$0.01  &   12.0   \\
GLMIC{\_}l343  & [343,344]     & [-1,+1]     & 0.639$\pm$0.001 & 0.622$\pm$0.001 & 0.516$\pm$0.001 & 0.548$\pm$0.001 & 0.60$\pm$0.01 & 0.47$\pm$0.01 & 0.40$\pm$0.01 & 0.52$\pm$0.01  &   12.0   \\
GLMIC{\_}l344  & [344,345]     & [-1,+1]     & 0.642$\pm$0.001 & 0.635$\pm$0.001 & 0.527$\pm$0.001 & 0.558$\pm$0.001 & 0.61$\pm$0.01 & 0.50$\pm$0.01 & 0.43$\pm$0.01 & 0.55$\pm$0.01  &   12.0   \\
GLMIC{\_}l345  & [345,346]     & [-1,+1]     & 0.639$\pm$0.001 & 0.611$\pm$0.001 & 0.506$\pm$0.001 & 0.533$\pm$0.001 & 0.61$\pm$0.01 & 0.49$\pm$0.01 & 0.46$\pm$0.01 & 0.59$\pm$0.01  &   12.0   \\
GLMIC{\_}l346  & [346,347]     & [-1,+1]     & 0.654$\pm$0.001 & 0.642$\pm$0.001 & 0.529$\pm$0.001 & 0.559$\pm$0.001 & 0.64$\pm$0.01 & 0.58$\pm$0.01 & 0.49$\pm$0.01 & 0.60$\pm$0.01  &   12.0   \\
GLMIC{\_}l347  & [347,348]     & [-1,+1]     & 0.655$\pm$0.001 & 0.657$\pm$0.001 & 0.533$\pm$0.001 & 0.570$\pm$0.001 & 0.63$\pm$0.01 & 0.58$\pm$0.01 & 0.50$\pm$0.01 & 0.63$\pm$0.01  &   12.0   \\
GLMIC{\_}l348  & [348,349]     & [-1,+1]     & 0.659$\pm$0.001 & 0.652$\pm$0.001 & 0.531$\pm$0.001 & 0.561$\pm$0.001 & 0.62$\pm$0.01 & 0.53$\pm$0.01 & 0.47$\pm$0.01 & 0.61$\pm$0.01  &   12.0   \\
GLMIC{\_}l349  & [349,350]     & [-1,+1]     & 0.649$\pm$0.001 & 0.649$\pm$0.001 & 0.527$\pm$0.001 & 0.554$\pm$0.001 & 0.60$\pm$0.01 & 0.51$\pm$0.01 & 0.44$\pm$0.01 & 0.59$\pm$0.01  &   12.0   \\
GLMXC{\_}l350  & [350,351]     & [-1,+1]     & 0.651$\pm$0.001 & 0.645$\pm$0.001 & 0.525$\pm$0.001 & 0.544$\pm$0.001 & 0.60$\pm$0.01 & 0.49$\pm$0.01 & 0.47$\pm$0.01 & 0.58$\pm$0.01  &   11.5   \\
GLMIIC{\_}l351 & [351,352]     & [-1,+1]     & 0.646$\pm$0.001 & 0.635$\pm$0.001 & 0.517$\pm$0.001 & 0.539$\pm$0.001 & 0.59$\pm$0.01 & 0.49$\pm$0.01 & 0.45$\pm$0.01 & 0.56$\pm$0.01  &   11.5   \\
GLMIIC{\_}l352 & [352,353]     & [-1,+1]     & 0.644$\pm$0.001 & 0.623$\pm$0.001 & 0.512$\pm$0.001 & 0.536$\pm$0.001 & 0.60$\pm$0.01 & 0.49$\pm$0.01 & 0.48$\pm$0.01 & 0.58$\pm$0.01  &   11.5   \\
GLMIIC{\_}l353 & [353,354]     & [-1,+1]     & 0.646$\pm$0.001 & 0.620$\pm$0.001 & 0.514$\pm$0.001 & 0.536$\pm$0.001 & 0.60$\pm$0.01 & 0.59$\pm$0.01 & 0.51$\pm$0.01 & 0.65$\pm$0.01  &   12.0   \\
GLMIIC{\_}l354 & [354,355]     & [-1,+1]     & 0.665$\pm$0.001 & 0.637$\pm$0.001 & 0.531$\pm$0.001 & 0.555$\pm$0.001 & 0.61$\pm$0.01 & 0.63$\pm$0.01 & 0.52$\pm$0.01 & 0.67$\pm$0.01  &   12.0   \\
GLMIIC{\_}l355 & [355,356]     & [-1.5,+1.5] & 0.651$\pm$0.001 & 0.621$\pm$0.001 & 0.519$\pm$0.001 & 0.547$\pm$0.001 & 0.61$\pm$0.01 & 0.60$\pm$0.01 & 0.52$\pm$0.01 & 0.66$\pm$0.01  &   11.5   \\
GLMIIC{\_}l356 & [356,357]     & [-1.5,+1.5] & 0.653$\pm$0.001 & 0.612$\pm$0.001 & 0.520$\pm$0.001 & 0.549$\pm$0.001 & 0.64$\pm$0.01 & 0.61$\pm$0.01 & 0.51$\pm$0.02 & 0.70$\pm$0.01  &   11.5   \\
GLMIIC{\_}l357 & [357,358]     & [-1.5,+1.5] & 0.643$\pm$0.001 & 0.593$\pm$0.001 & 0.510$\pm$0.001 & 0.538$\pm$0.001 & 0.63$\pm$0.01 & 0.68$\pm$0.02 & 0.49$\pm$0.02 & 0.71$\pm$0.02  &   11.5   \\
GLMIIC{\_}l358 & [358,359]     & [-2,+2]     & 0.632$\pm$0.001 & 0.582$\pm$0.001 & 0.499$\pm$0.001 & 0.527$\pm$0.001 & 0.63$\pm$0.01 & 0.70$\pm$0.01 & 0.53$\pm$0.01 & 0.72$\pm$0.01  &   11.5   \\
GLMIIC{\_}l359 & [359,360]     & [-2,+2]     & 0.630$\pm$0.001 & 0.580$\pm$0.001 & 0.498$\pm$0.001 & 0.522$\pm$0.001 & 0.63$\pm$0.01 & 0.68$\pm$0.01 & 0.49$\pm$0.02 & 0.72$\pm$0.01  &   11.5   \\
\enddata
\tablenotetext{a} {{\footnotesize GLMXC{\_}l009, GLMXC{\_}l350,
GLMIC{\_}l064 and GLMIC{\_}l295 are the combined fields (see
\S\ref{DATA}).}}
\end{deluxetable}

\begin{table}
\begin{center}
\caption{Dispersion ranges of the extinction ratios
$A_{\lambda}/A_{\Ks}$ derived from red giants (RG1).
\label{tbl-extrange}} \vspace{0.2in}
\begin{tabular}{c|ccc}
\tableline\tableline
$\lambda$ ($\mum$) & Minimum & Maximum & range \\
\tableline 3.6 (3.545) & 0.61 & 0.68 & 0.07 \\
4.5 (4.442)& 0.57 & 0.73 & 0.16 \\
5.8 (5.675)& 0.48 & 0.62 & 0.14 \\
8.0 (7.760)& 0.47 & 0.59 & 0.12 \\
\tableline 7.0\tablenotemark{a} & 0.37 & 0.55 & 0.18 \\
15.0 & 0.19 & 0.54 & 0.35 \\
\tableline
\end{tabular}
\tablenotetext{a}{The 7$\mum$ and 15$\mum$ extinction data are taken
from \citet{Jiang06}.}
\end{center}
\end{table}

\end{document}